# Penetration of boundary-driven flows into a rotating spherical thermally-stratified fluid

**Grace A. Cox[1,2]†, Christopher J. Davies[3], Philip W. Livermore[3], and James Singleton[3]**

[1]School of Environmental Sciences, University of Liverpool, Liverpool, L69 3GP, UK

[2]Department of Applied Mathematics, University of Leeds, Leeds, LS2 9JT, UK

[3]School of Earth and Environment, University of Leeds, Leeds, LS2 9JT, UK

Motivated by the dynamics within terrestrial bodies, we consider a rotating, strongly thermally stratified fluid within a spherical shell subject to a prescribed laterally inhomogeneous heat-flux condition at the outer boundary. Using a numerical model, we explore a broad range of three key dimensionless numbers: a thermal stratification parameter (the relative size of boundary temperature gradients to imposed vertical temperature gradients), $10^{-3} \leqslant S \leqslant 10^4$, a buoyancy parameter (the strength of applied boundary heat flux anomalies), $10^{-3} \leqslant B \leqslant 10^6$, and the Ekman number (ratio of viscous to Coriolis forces), $10^{-6} \leqslant E \leqslant 10^{-4}$. We find both steady and time-dependent solutions and delineate the temporal regime boundaries. We focus on steady-state solutions, for which a clear transition is found between a low $S$ regime, in which buoyancy dominates dynamics, and a high $S$ regime, in which stratification dominates. For the latter case, the radial and horizontal velocities scale respectively as $u_r \sim S^{-1}$, $u_h \sim S^{-\frac{3}{4}}\, B^{\frac{1}{4}}$ and are confined to boundary-induced flow within a thin layer of depth $(S\, B)^{-\frac{1}{4}}$ at the outer edge of the domain. For the Earth, if lower-mantle heterogeneous structure is due principally to chemical anomalies, we estimate that the core is in the high-$S$ regime and steady flows arising from strong outer-boundary thermal anomalies cannot penetrate the stable layer. However, if the mantle heterogeneities are due to thermal anomalies and the heat-flux variation is large, the core will be in a low-$S$ regime in which the stable layer is likely penetrated by boundary-driven flows.

† Email address for correspondence: gracecox@cp.dias.ie

## 1. Introduction

Differential heating at the boundary of a stratified fluid arises in a variety of physical systems. The oceans and atmosphere are heated non-uniformly from above owing to the latitudinal variation of incoming solar energy. Fluid near the differentially heated surface moves laterally away from anomalously warm regions towards anomalously cold regions and a significant amount of work has considered whether this 'horizontal convection' can drive large-scale overturning circulations (e.g. Paparella & Young 2002; Siggers *et al.* 2004; Sheard *et al.* 2016; Shishkina 2017). The primary motivation for the present study is differential heating of planetary cores due to lateral heat flow anomalies in their overlying solid mantles. We conduct a systematic investigation of the interaction between thermal stratification and differential boundary heating, incorporating the key ingredients of rapid rotation and spherical shell geometry. Our main focus is to establish the extent to which boundary heat flow anomalies can penetrate and disrupt a pre-existing thermal stratification.

There is now a body of evidence indicating that the cores of Mercury (Christensen 2006), Earth (Davies *et al.* 2015; Nimmo 2015), Mars (Stevenson 2001) and Ganymede (Rückriemen *et al.* 2015) are thermally stably stratified below the core-mantle boundary (CMB) owing to a subadiabatic CMB heat flow, with convection (and magnetic field generation) arising at greater depths. The existence of stratification is important because it influences the intensity and structure of the observable magnetic field (Christensen 2006; Stanley & Glatzmaier 2010) and reflects the core's long-term evolution. The strength and thickness of these thermally stable regions is hard to assess due to a lack of direct observations. The stable layer in Earth's core could be up to $\sim$700 km thick (Gubbins *et al.* 2015) with a Brunt-Väisälä frequency comparable to the rotation period. Thermal stratification in the Martian core is usually estimated to have begun around 4 Ga, corresponding to the epoch when the planet lost its global magnetic field (Stevenson 2001), and so the thermally stable region could occupy a significant fraction of the present-day core. Thermal history models for Ganymede predict a stable layer hundreds of kilometres thick (Rückriemen *et al.* 2015).

Terrestrial planetary cores are overlain by rocky mantle, which acts like a viscous fluid convecting on timescales of $10^8$ years. In contrast, liquid metal cores have very low viscosity and convect on timescales of $10^3$ years. This difference in convection timescales means that the core responds to the CMB as a rigid surface with a fixed heat flux imposed by the lower mantle, whilst the mantle is subjected to a uniform temperature lower boundary condition (Olson & Christensen 2002). Mantle convection simulations produce lateral temperature anomalies of thousands of Kelvin and lateral CMB heat flow variations greater than the mean CMB heat flow (e.g. Nakagawa & Tackley 2008; Olson *et al.* 2015). These lateral variations will inevitably drive baroclinic flows in the underlying core through the thermal wind, but it is unclear the extent to which they will drive penetrative flow within a strongly stratified region.

The competition between stratification and boundary forcing has been explored in some numerical studies of convection in nonmagnetic rotating spherical shells, which have shown that thermal boundary anomalies are capable of drastically altering the dynamics compared to uniform thermal boundary conditions (e.g. Zhang & Gubbins 1992, 1993; Gibbons & Gubbins 2000; Gibbons *et al.* 2007). Zhang & Gubbins (1992) solved for steady flows driven by lateral thermal variations at the outer boundary of a rotating spherical shell, having specified temperature rather than heat flux for numerical simplicity. They studied both unstratified and weakly stratified fluids subjected to a range of temperature anomaly patterns and magnitudes. For modest boundary anomaly strengths, patterns of temperature fluctuations and fluid flow lock to the boundary anomaly pattern through the thermal wind, and flows penetrate deep into the shell due to Coriolis effects. Stratification greatly reduces radial flow amplitudes, though toroidal flows are less affected, and confine flow towards the outer boundary. The authors speculated that these results would also be obtained in the geophysical case of fixed heat flux boundary anomalies. Gibbons & Gubbins (2000) were able to confirm this for steady flows in their subsequent investigation of weakly stratified fluids in rotating spherical shells. They applied different spatial distributions and magnitudes of large-scale boundary heat flow anomalies to fluids of varying stratification strengths. For equatorially symmetric patterns, rotational effects dominate dynamics at weak or no stratification. As the stratification increases, rotational effects become less important, radial flow diminishes and flow is confined to a layer beneath the outer boundary. Smaller length scale heat flux patterns drive less energetic flows that are not able to penetrate as deeply into the fluid. Solutions become increasingly smaller scale with increasing boundary anomaly magnitude, with correspondingly higher computational expense. Gibbons & Gubbins (2000) suggested that solutions would become unstable (time-dependent) with sufficiently strong boundary anomalies, though computational limitations prevented the authors from identifying the parameters at which this occurs.

Several authors have considered the more realistic but more complex magnetohydrodynamic (MHD) case by studying numerical simulations of dynamos in partially stratified spherical shells, including Christensen (2006); Christensen & Wicht (2008);

Stanley & Mohammadi (2008); Aurnou & Aubert (2011); Nakagawa (2011, 2015); Olson *et al.* (2017). Some numerical models have shown that the presence of a stable layer fundamentally changes dynamo action and can drastically alter the magnetic field at the planetary surface compared to equivalent models with no stable layer. For example, Christensen (2006) showed that a strong magnetic field at the top of the dynamo generating region diffuses through a stable layer such that the small-scale, rapidly varying components are filtered out.

Dynamo models with heterogeneous thermal boundary conditions have also been investigated by various authors, see the review by Amit *et al.* (2015) and references therein. As in the non-magnetic case, within MHD models heterogeneous boundary forcing has been shown to have a significant effect, for example by modifying the morphology of the magnetic field (e.g. Olson & Christensen 2002; Gubbins *et al.* 2007; Aurnou & Aubert 2011) such that its long-term fundamental symmetries follow the spatial symmetries of the imposed heat flux pattern, or by locking the magnetic field to regions of anomalously high heat flow (Willis *et al.* 2007; Sreenivasan 2009). In some circumstances, strong boundary driven flows can also overwhelm the convection such that dynamo action is weakened or destroyed altogether (Olson & Christensen 2002; Takahashi *et al.* 2008), though this is not necessarily the case (Aurnou & Aubert 2011). Although ultimately the most physically relevant model, a thorough scaling analysis of the competition between stratification and boundary forcing within an MHD setting is beyond what is currently achievable. Some progress has been made by studying weakly stratified models with heterogeneous outer boundary conditions (e.g. Sreenivasan & Gubbins 2008; Aurnou & Aubert 2011; Olson *et al.* 2017), although the extrapolation gap from the parameters used in these models to realistic values is large.

This work focusses on the simpler, non-magnetic problem which is yet not fully described. In particular, the previous studies described above have been limited to highly viscous, weakly-stratified fluids in spherical shells with moderate rotation rates and subject to relatively weak boundary anomalies: it is not clear how these results bear on the rapidly rotating, strongly stratified case relevant to planetary cores that are additionally subject to significant lateral variations in heat flux at their outer boundary. One severe computational limitation that has hampered progress arises because rotating flows adopt small azimuthal length scales even at the onset of convection (Chandrasekhar 1961), while increasing the amplitude of the driving force generates a broad spectrum of flow structures that become increasingly difficult to resolve. In this study, we minimise this problem by considering a subset of steady-state solutions obtained from solving the full time-dependent equations, and also by assuming that the entire fluid domain is stably stratified without any internal heat sources that drive internal convection. This is equivalent to assuming that any underlying convection does not significantly penetrate or mix an overlying stable region, which is true in the case of strong stratification (Takehiro & Lister 2001; Buffett & Seagle 2010; Gubbins & Davies 2013). These assumptions allow us to isolate the interaction between outer-boundary forcing and pre-existing stratification, without the additional complication of destabilisation of stratified fluid from below by internal convection, and to study the dynamics using a much wider range of parameters than has been possible previously.

The fluid dynamical problem we consider depends upon three dimensionless numbers (detailed definitions are given in 2.1): a thermal stratification parameter, $S$, defined as the relative size of boundary temperature gradients to imposed vertical temperature gradients, a buoyancy parameter, $B$, measuring the strength of the applied boundary heat flux anomalies, and the Ekman number, $E$, the ratio of viscous and Coriolis forces. Our study spans the ranges $10^{-3} \leqslant B \leqslant 10^{6}$, $10^{-3} \leqslant S \leqslant 10^{4}$ and $10^{-6} \leqslant E \leqslant 10^{-4}$. We focus primarily on the case where the aspect ratio, the ratio of inner to outer boundary radii, corresponds to that of Earth's liquid core, $r_i/r_o = \eta = 0.35$. Additional simulations are performed at $\eta = 0.01$, which is almost a full sphere and approximates the core geometry of Mars and Ganymede.

For each choice of $(E, S, B)$ a heat flow pattern must be chosen. Previous studies clearly show that the influence of thermal boundary anomalies on the structure and dynamics of rotating fluids becomes more pronounced as the lengthscale of the imposed pattern is increased (Zhang & Gubbins 1992, 1993; Davies *et al.* 2009). We choose to apply a $Y_2^2$ spherical harmonic boundary heat flow pattern since this the largest component of shear wave variation (a likely proxy for CMB heat flow) in Earth's lower mantle (Dziewonski *et al.* 2010); it is also a common boundary condition of previous studies, which makes comparison straightforward (e.g. Zhang & Gubbins 1992, 1993; Davies *et al.* 2009; Sreenivasan 2009; Sahoo & Sreenivasan 2017).

We have conducted a suite of 110 numerical simulations finding predominantly steady solutions, which partition into two distinct regimes. Within each regime we formulate theoretical scaling laws that provide excellent fits to our dataset and permit extrapolation to the parameter regimes appropriate to planetary interiors. The remainder of the paper is structured as follows: the mathematical formulation is given in §2, results of the numerical simulations are presented in §3, scaling analyses and their application to Earth and Ganymede's outer cores follow in §4 and §5, and a summary of results is found in §6.

## 2. Method

We consider an incompressible Boussinesq fluid in an impenetrable spherical shell, of outer radius $r_o$ and inner radius $r_i$, rotating about the axial $\hat{\boldsymbol{z}}$ direction with constant angular velocity $\Omega$. The whole shell is thermally stratified and compositional effects are neglected, as in Gibbons & Gubbins (2000), in order to isolate the effects of thermal boundary anomalies on a thermally stratified fluid. Again following Gibbons & Gubbins (2000), we also neglect the magnetic field so as to reach more realistic $E$, $B$ and $S$ values; the effects of free convection and the resulting magnetic field evolution will be investigated in a future study. In the following work, $r$, $\theta$ and $\phi$ denote spherical polar coordinates, $\boldsymbol{r}$ is the position vector and $t$ is time.

### 2.1. *Governing equations and non-dimensionalisation*

Following the formulation of Zhang & Gubbins (1992) and Gibbons & Gubbins (2000), the temperature is split into a steady radial part, $T_0$, and a time-varying part, $T_1$, such that

$$T(r,\theta,\phi,t) = T_0(r) + T_1(r,\theta,\phi,t). \tag{2.1}$$

The steady radial temperature profile satisfies

$$\kappa\nabla^2 T_0 = F, \tag{2.2}$$

where $\kappa$ is the thermal diffusivity and $F \geqslant 0$ is a heat sink, and is chosen to impose a background thermal gradient that, if strong, suppresses radial motion. Integrating with respect to $r$ in spherical coordinates gives

$$r^2\frac{\mathrm{d}T_0}{\mathrm{d}r} = \beta r^3 + A \tag{2.3}$$

where $\beta = \frac{F}{3\kappa}$ and $A$ is a constant of integration. Setting the outer boundary condition such that

$$\left.\frac{\mathrm{d}T_0}{\mathrm{d}r}\right|_{r=r_o} = \beta r_o \tag{2.4}$$

results in $A = 0$ and so within the spherical shell $\frac{\mathrm{d}T_0}{\mathrm{d}r} = \beta r$.

We define the outer boundary condition of the temperature gradient as

$$\left.\frac{\partial T_1}{\partial r}\right|_{r=r_o} = \mathcal{H} Y_2^2(\theta, \phi), \tag{2.5}$$

in which the spatial pattern of the anomaly is given by the spherical harmonic $Y_2^2(\theta, \phi)$, and the magnitude of the anomaly is given by $\mathcal{H}$. Rewriting the general temperature equation

$$\frac{\partial T}{\partial t} + (\boldsymbol{u} \cdot \nabla)T = \kappa \nabla^2 T - F, \tag{2.6}$$

using (2.1) and (2.4) leaves

$$\frac{\partial T_1}{\partial t} + (\boldsymbol{u} \cdot \nabla)T_1 + u_r \beta r = \kappa \nabla^2 T_1 \tag{2.7}$$

as the relevant temperature equation.

The equations for conservation of momentum in a rotating frame of reference and for conservation of mass are

$$\frac{\partial \boldsymbol{u}}{\partial t} + (\boldsymbol{u} \cdot \nabla)\boldsymbol{u} + 2\Omega(\hat{\boldsymbol{z}} \times \boldsymbol{u}) = -\nabla \left( \frac{P'}{\rho_0} \right) + \frac{\rho' \boldsymbol{g}}{\rho_0} + \nu \nabla^2 \boldsymbol{u} \tag{2.8}$$

and

$$\nabla \cdot \boldsymbol{u} = 0 \tag{2.9}$$

where $\boldsymbol{u}$ is velocity, $P'$ is the pressure perturbation, $\rho_0$ is a reference density, $\rho'$ is the deviation from the reference density, $\boldsymbol{g}$ is gravity and $\nu$ is the kinematic viscosity. Expressing $\rho'$ as

$$\rho' = -\rho_0 \alpha_T T, \tag{2.10}$$

where $\alpha_T$ is the coefficient of thermal expansivity, gives an alternative form of the

momentum equation

$$\frac{\partial \boldsymbol{u}}{\partial t} + (\boldsymbol{u} \cdot \nabla)\boldsymbol{u} + 2\Omega(\hat{\boldsymbol{z}} \times \boldsymbol{u}) = -\nabla \hat{P} + \alpha_T \gamma T \boldsymbol{r} + \nu \nabla^2 \boldsymbol{u}, \tag{2.11}$$

where $\hat{P}$ is the reduced pressure ($= P'/\rho_0$) and $\gamma$ is a constant ($\boldsymbol{g} = -\gamma \boldsymbol{r}$).

Scaling radius by a characteristic length scale $d$ ($= r_o - r_i$), time by the thermal diffusion time $d^2/\kappa$, velocity by $\kappa/d$ and temperature by $\mathcal{H}d$ (from equation (2.5)) gives the radial temperature profile and the temperature and momentum equations in their dimensionless forms

$$\frac{\mathrm{d}T_0^*}{\mathrm{d}r^*} = S\ r^*, \tag{2.12}$$

$$\frac{\partial T_1^*}{\partial t^*} + (\boldsymbol{u}^* \cdot \nabla)\, T_1^* + S\ u_r^* r^* = \nabla^2 T_1^* \tag{2.13}$$

and

$$\frac{E}{Pr}\left[\frac{\partial \boldsymbol{u}^*}{\partial t^*} + (\boldsymbol{u}^* \cdot \nabla)\boldsymbol{u}^*\right] + (\hat{\boldsymbol{z}}^* \times \boldsymbol{u}^*) = -\nabla \hat{P} + B\ T^* \boldsymbol{r}^* + E \nabla^2 \boldsymbol{u}^*, \tag{2.14}$$

where $\boldsymbol{r}^*$ is the dimensionless radial vector, $S$ is the stratification parameter, $E$ is the Ekman number, $Pr$ is the Prandtl number and $B$ is the buoyancy parameter. These dimensionless numbers are defined as

$$S = \frac{\beta d}{\mathcal{H}}, E = \frac{\nu}{2\Omega d^2}, Pr = \frac{\nu}{\kappa}, B = \frac{\alpha_T \gamma \mathcal{H} d^3}{2\Omega \kappa}, \tag{2.15}$$

and $B$ is related to $E$ and a Rayleigh number, $Ra_{\mathcal{H}}$, where

$$\frac{B}{E} = Ra_{\mathcal{H}} = \frac{\alpha_T \gamma \mathcal{H} d^5}{\nu \kappa}. \tag{2.16}$$

In this work, all calculations are performed at $Pr = 1$ for numerical convenience and the majority with a shell aspect ratio $\eta = 0.35$; a summary of model parameters is given in tables A.1 to A.4 in appendix A. We investigate the effects of varying the shell aspect ratio using models with $\eta = 0.01$ in section 4.3. The governing equations are solved for $\boldsymbol{u}$ and $T_1$ with no-slip boundary conditions on both inner and outer boundaries, a fixed temperature imposed on the inner boundary, and a fixed heat flux imposed on the outer boundary as previously discussed. A detailed description of the pseudo-spectral code may be found in Willis *et al.* (2007) and Davies *et al.* (2011), and in the most recent dynamo benchmark paper (Matsui *et al.* 2016). Although equations (2.1) – (2.5) give the clearest mathematical description of our method, in fact the code solves the following equation

$$\frac{\partial T^*}{\partial t^*} + (\boldsymbol{u}^* \cdot \nabla)\, T^* = \nabla^2 T^* - 3\ S, \tag{2.17}$$

which is equivalent to (2.13). To benchmark our code for this particular problem, we reproduced the flow magnitudes and spatial patterns reported in Gibbons & Gubbins (2000), using a shell aspect ratio $\eta = 0.4$ and their parameters of $E = 10^{-3}$, $Pr = 1$, $B = 1$ and $S = 0$ and $S = 100$.

Given that we focus upon steady-state solutions to the time-dependent equations, for numerical expediency where possible we used the final steady-state solution of a model nearby in parameter space as the initial condition. Models were run long past the initial transient period and until the volume-averaged kinetic energy converged to a steady value. Several numerical models were unstable and no steady-state solutions were obtained at those parameters. In such cases, we cannot rule out the existence of a steady-state model using different initial conditions.

For each of our 110 models, spatial convergence was verified by assessing the kinetic energy power spectrum as a function of spherical harmonic degree ($l$) and order ($m$). For all models, the maximum power was found at long wavelengths (the lowest $l$), which generally exceeded the power in the shortest wavelengths (high $l$) by a large amount: at least two, though usually four or five, orders of magnitude.

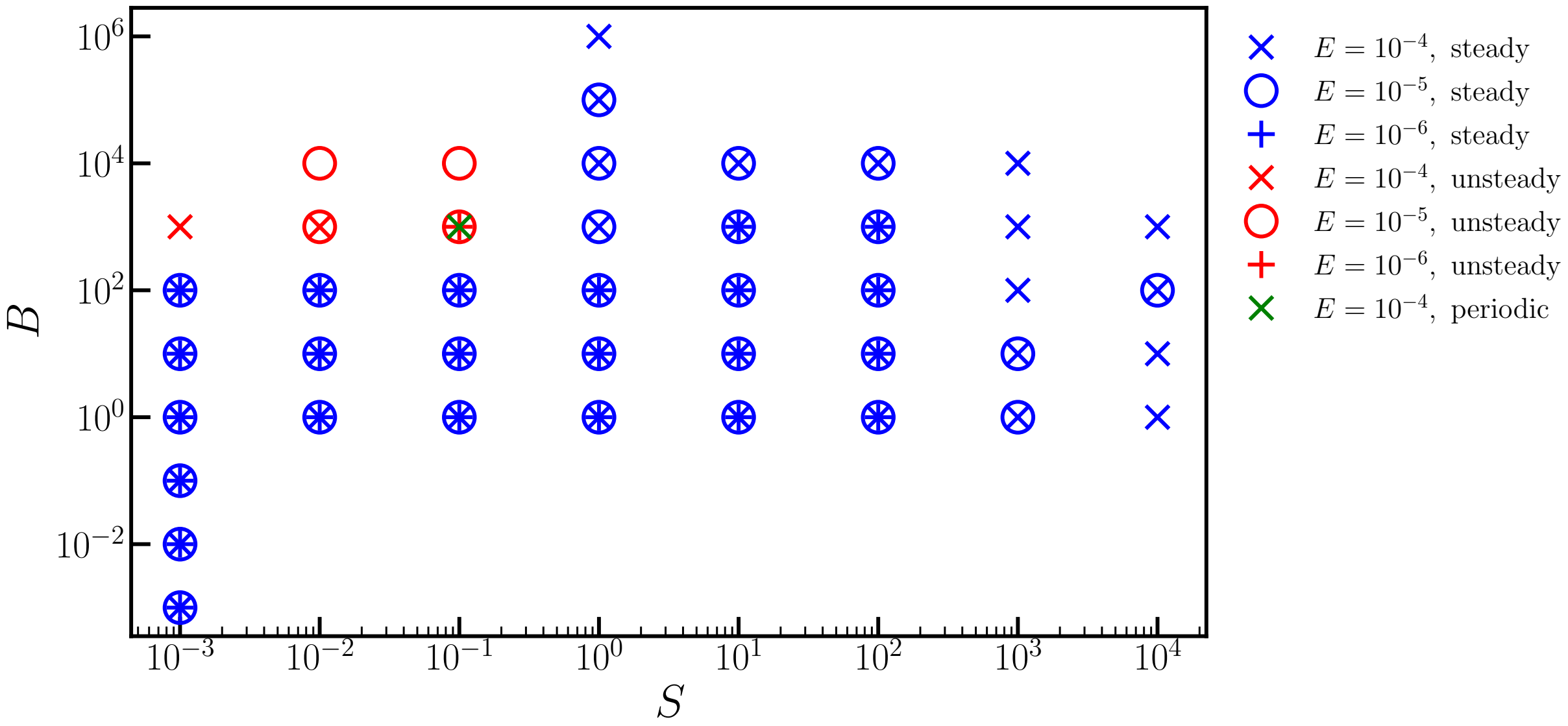

Figure 1: Stability diagram in ($S$, $B$) parameter space showing all models summarised in tables A.1 to A.3. The symbol type represents the Ekman number (crosses denote $E = 10^{-4}$, circles denote $E = 10^{-5}$ and plus signs denote $E = 10^{-6}$); the symbol colour represents the stability of the solution obtained (blue denotes a steady state solution, red denotes a time dependent solution and green denotes a periodic solution).

Fig. 1 is a stability diagram showing regions of parameter space resulting in steady and unsteady solutions. The figure shows the transition between high $B$ and low $S$ models, which are unsteady, and higher $S$ models, which produce a steady state. One periodic model was obtained at the boundary between the steady and unsteady regions of parameter space. In the remainder of this work, we focus our attention upon the steady-state regime; time-dependent models are the subject of a future paper.

## 3. Results

Fig. 2 shows the temperature perturbations, $T_1^*$, in the equatorial plane for models at $E = 10^{-4}$ and a range of $B$ and $S$ values. Figs 3 and 4 show the radial and azimuthal velocity components, $u_r^*$ and $u_\phi^*$, for the same models. At low $B$ and $S$, the temperature fluctuations are large-scale with a $Y_2^2$ spatial pattern locked to the applied heat flux pattern on the outer boundary and penetrating through the whole shell depth. The two lobes of negative temperature (blue) correspond to regions of high outward heat flux and the two lobes of positive temperature (red) correspond to regions of low outward heat flux. Zeroes of $T_1^*$ (at $\phi \approx \pi/4, 3\pi/4, 5\pi/4, 7\pi/4$) correspond to locations of the outer boundary heat flux changing sign. The radial velocity is dominated by large-scale convection cells that occupy the whole shell, with two upwellings and two downwellings present, and the peak velocity amplitudes occur at approximately half the shell radius. The lateral locations of these maxima and minima approximately correspond to locations of $T_1^* = 0$. In azimuthal velocity, locations of diverging (converging) lobes of opposite sign correspond to locations of upwellings (downwellings) of radial flow and $T_1^* = 0$.

As the stratification parameter ($S$) increases, temperature perturbations and flow magnitudes decrease and the dynamics become concentrated towards the outer boundary rather than occupying the entire shell thickness. Radial flow cells begin to elongate near the inner boundary, and high velocity magnitudes are concentrated near the outer boundary rather than the inner boundary. In $u_\phi^*$, inner and outer cells of the same polarity begin to join together through tails trailing from the outermost cells, with the inner cells decreasing in amplitude. Radial flow is strongly suppressed with increasing $S$, which is expected because stratification does not permit large radial velocities. Azimuthal flow is only weakly suppressed with increasing stratification as horizontal flows are permitted within a stably stratified layer. At high $S$, all flow becomes confined to a thin shear layer of thickness $\delta^*$ beneath the outer boundary (hereafter referred to as the 'penetration depth' into the fluid).

As $B$ increases, temperature perturbations decrease and flow magnitudes increase. This is a consequence of the fixed heat flux outer boundary condition; increasing the buoyancy produces stronger flows that better homogenise the temperature, resulting in velocity increasing with $B$ while temperature perturbations decrease (e.g. Otero *et al.*

2002; Mound & Davies 2017). Flows are phase shifted so that upwellings (and diverging $u_\phi^*$) and downwellings (and converging $u_\phi^*$) are now locked to the boundary pattern itself rather to locations of heat flux changing sign. Upwellings (downwellings) are beneath high (low) boundary heat flow regions. At low $S$ and increasing $B$ (e.g. figs 2–4, a–c), temperature and flow patterns are strikingly different from models at other parameters. Downwellings become increasingly faster and much narrower in azimuth with increasing $B$, though still occupying the whole shell radius, whilst the upwellings remain broad and low amplitude. This pattern of slow, broad upwellings and fast, narrow downwellings in the presence of lateral boundary anomalies was also obtained in e.g. Willis *et al.* (2007); Sreenivasan & Gubbins (2011). At higher $S$, upwellings and downwellings are of similar lateral extent and dynamics are confined to a thin shear layer whose thickness decreases with increasing $S$ and $B$.

Fig. 5 shows $u_r^*$ (left) and $u_\phi^*$ (middle) and $T_1^*$ (right) in a meridional plane for models run at $E = 10^{-4}$ and $B = 1$ for a range of stratification parameters ($S$). At low $S$, dynamics are dominated by large-scale features that are aligned with the rotation axis. There is little variation parallel to the $z$-axis, as expected in a rapidly rotating system from the Taylor-Proudman theorem. As stratification increases, the dynamics are confined to the shear layer at the top of the shell, as seen in figures 2 to 4, which means that significant $z$ variations now occur in the models on the order of the penetration depth, $\delta^*$.

Fig. 6 shows $\langle u_r^* \rangle_v$, $\langle u_\phi^* \rangle_v$, $\langle v_\theta^* \rangle_v$ and $\langle T_1^* \rangle_v$, where the angular brackets denote the magnitude averaged over the shell volume $V$ such that, for example, $\langle u_r^* \rangle_v = \int |u_r^*| dV$, and likewise for vector quantities. We define a similar operator for the integral over a surface $S$ of radius $r$ such that $\langle u_r^* \rangle_s = \frac{1}{S} \int |u_r^*| dS$. We adopt an average over the entire domain, rather than only the shear layer volume, because it is difficult to estimate

(a) $B = 1, S = 0.1$ (b) $B = 10, S = 0.1$ (c) $B = 100, S = 0.1$

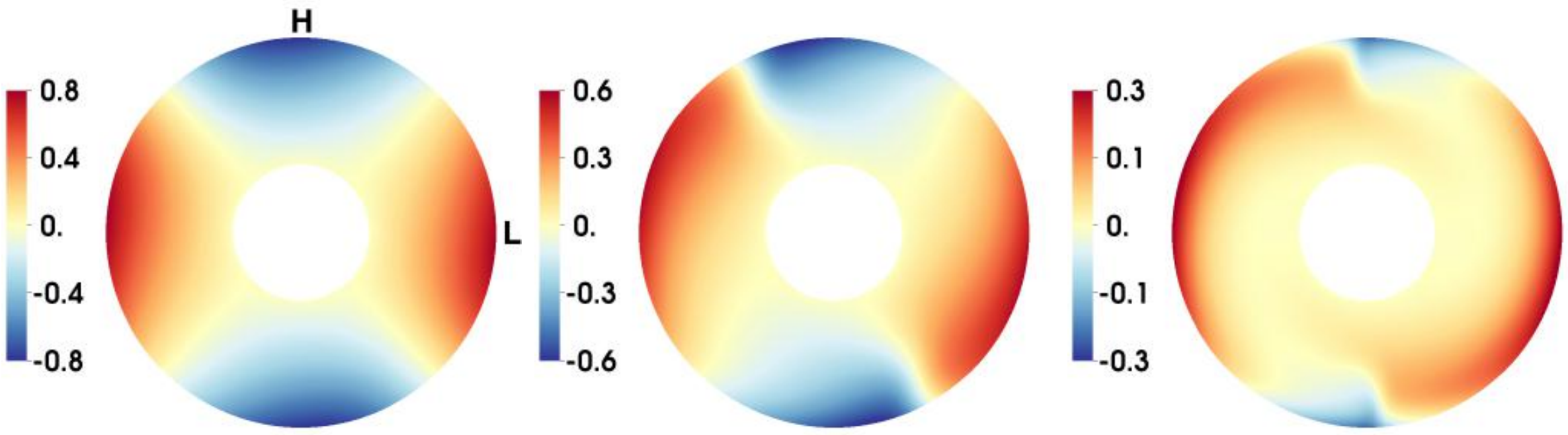

(d) $B = 1, S = 10$ (e) $B = 10, S = 10$ (f) $B = 100, S = 10$

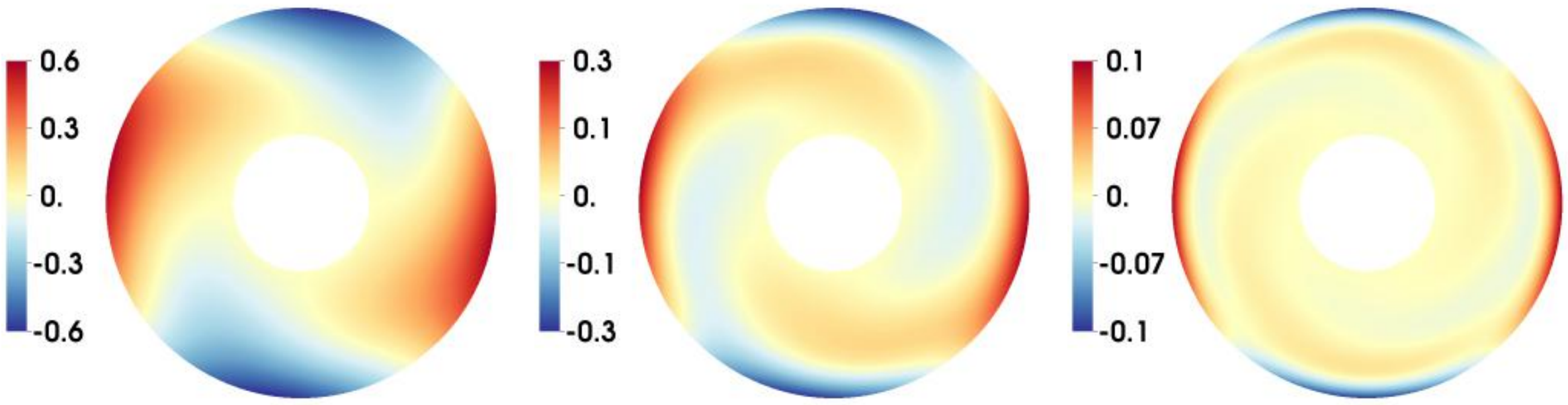

(g) $B = 1, S = 100$ (h) $B = 10, S = 100$ (i) $B = 100, S = 100$

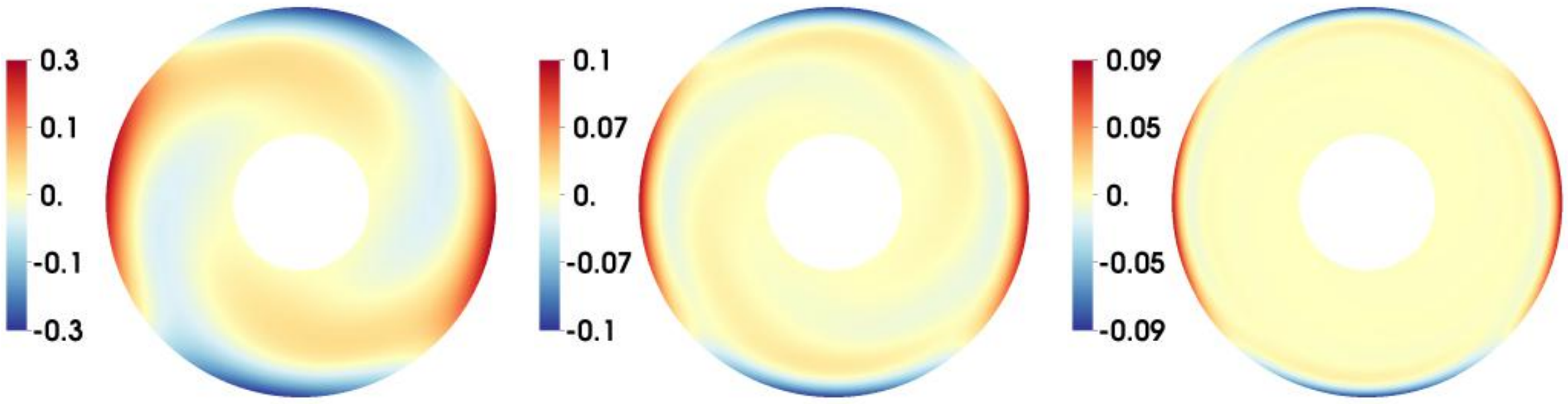

(j) $B = 1, S = 1000$ (k) $B = 10, S = 1000$ (l) $B = 100, S = 1000$

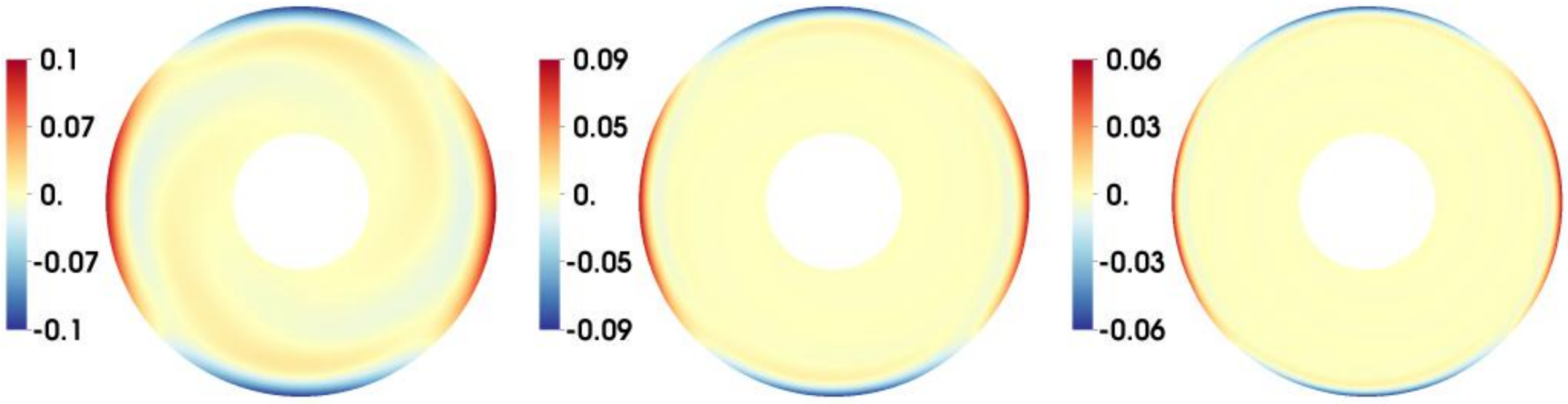

Figure 2: Equatorial plots of $T_1^*$ for models at $E = 10^{-4}$ and varying $S$ (increasing from top to bottom) and $B$ (increasing from left to right). Red indicates positive values and blue indicates negative values. Note the different colour scales. Locations of high (H) and low (L) outward heat flux are shown on the top left.

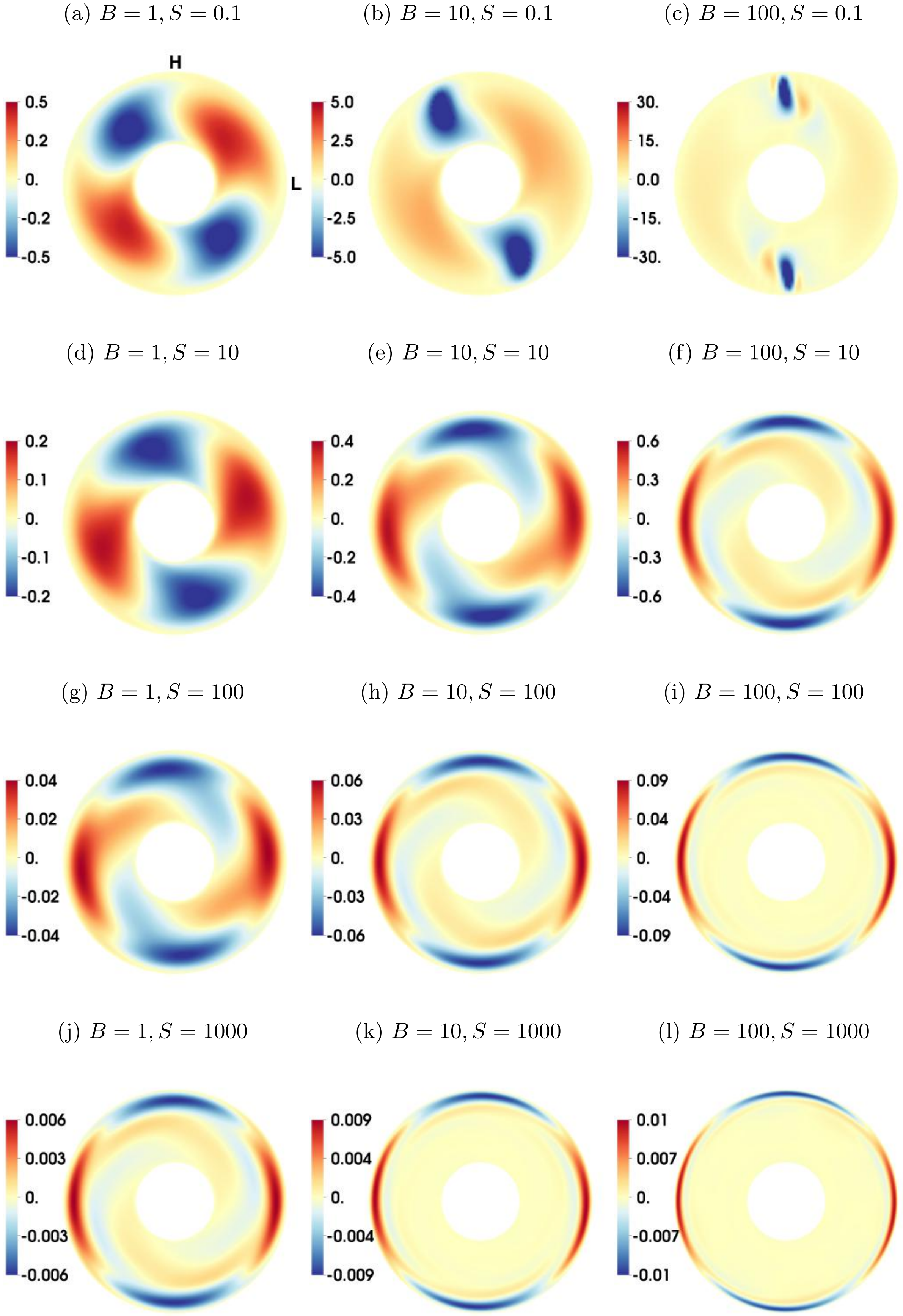

Figure 3: Equatorial plots of $u_r^*$ for models at $E = 10^{-4}$ and varying $S$ (increasing from top to bottom) and $B$ (increasing from left to right). Red indicates positive values and blue indicates negative values. Note the different colour scales. Locations of high (H) and low (L) outward heat flux are shown on the top left.

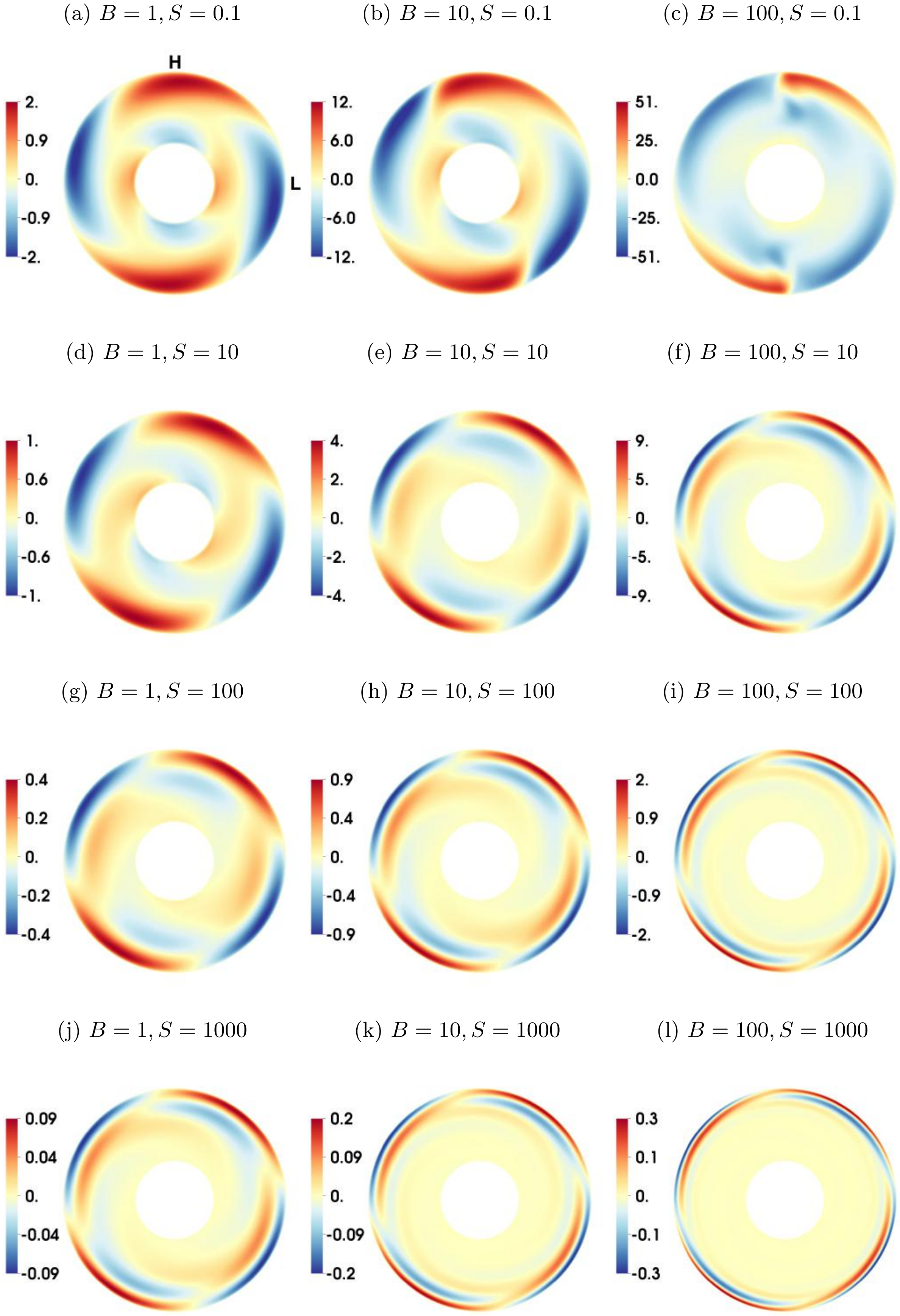

Figure 4: Equatorial plots of $u_\phi^*$ for models at $E = 10^{-4}$ and varying $S$ (increasing from top to bottom) and $B$ (increasing from left to right). Red indicates positive values and blue indicates negative values. Note the different colour scales. Locations of high (H) and low (L) outward heat flux are shown on the top left.

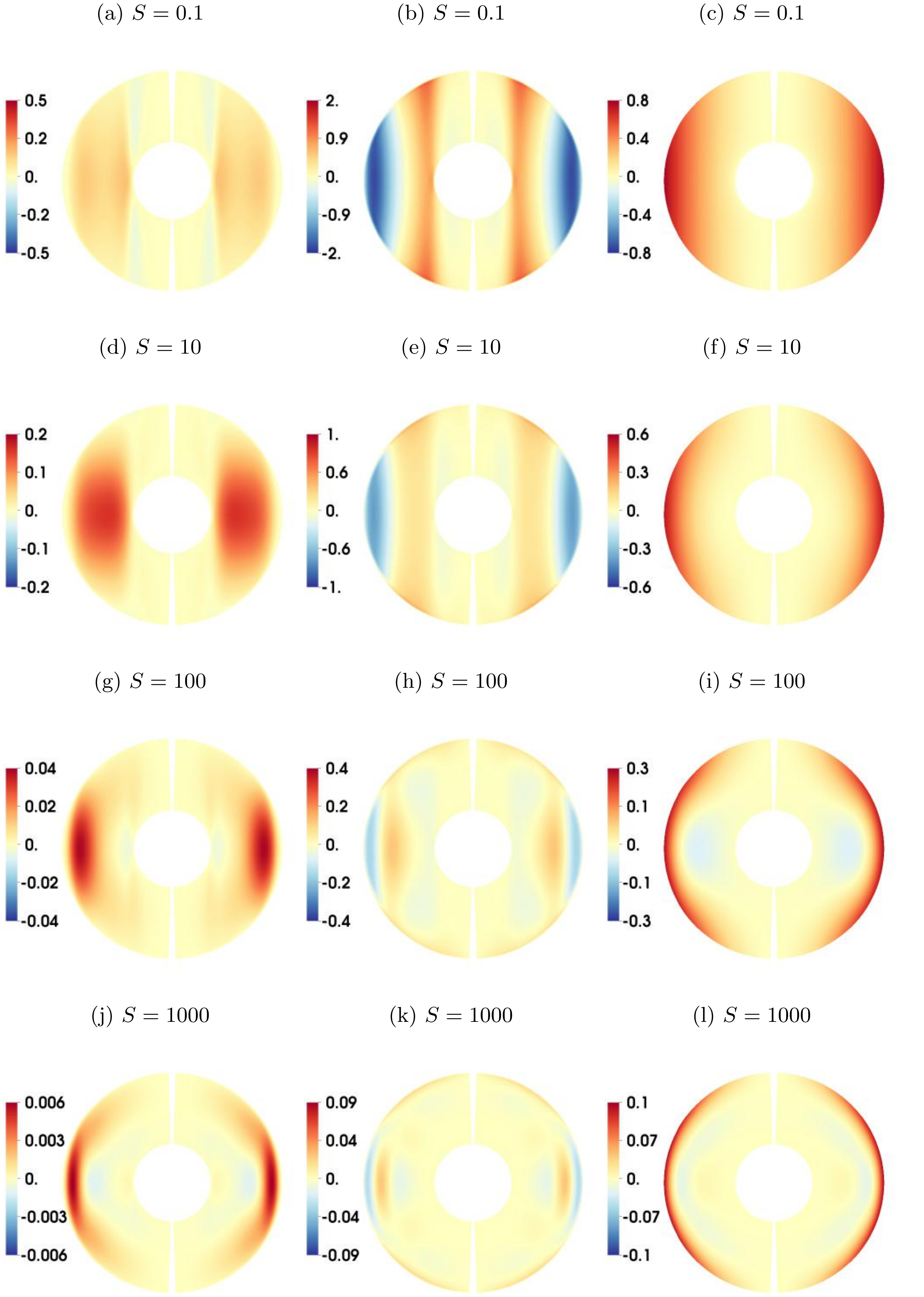

Figure 5: Meridional plots of $u_r^*$ (left), $u_\theta^*$ (middle) and $T_1^*$ (right) for models at $E = 10^{-4}$, $B = 1$ and varying $S$ (increasing from top to bottom). Red indicates positive values and blue indicates negative values.

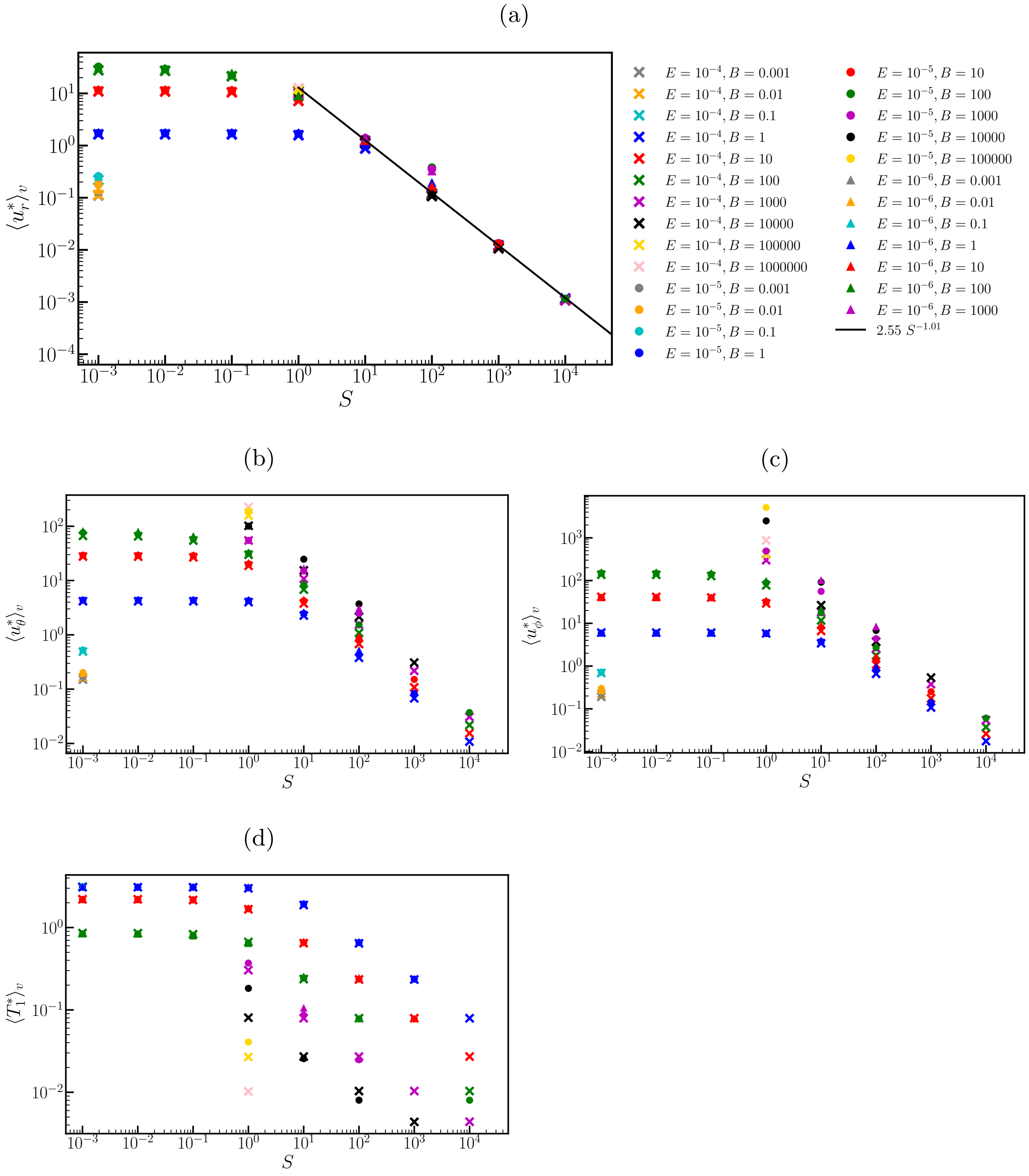

Figure 6: Volume-averaged values of the absolute (a) radial velocity, $\langle u_r^* \rangle_v$, (b) meridional velocity, $\langle u_\theta^* \rangle_v$, (c) azimuthal velocity, $\langle u_\phi^* \rangle_v$ and (d) temperature perturbations, $\langle T_1^* \rangle_v$, as a function of the stratification parameter, $S$, for all steady models. Symbol shapes represent the Ekman number, $E$, and colours represent the buoyancy parameter, $B$. The black line in panel (a) is the power law best fit for all models at $S > 1$.

the exact location of the shear layer edge. We assume that the quantities of interest are dominated by their values within the shear layer, with negligible contribution from elsewhere in the domain, such that our volume-averaged quantities are representative of the shear layer volume-average. Furthermore, we use the average of the modulus because integration over solid angle would otherwise result in large scale cancellation due to the spherical symmetry of the problem. The volume-averaged quantities show a clear transition from the low stratification ($S$) regime, in which dynamics appear to be related to $B$ and $E$ only, and the high $S$ regime, in which stratification dominates the dynamics and the quantities obey power law relationships in both $S$ and $B$.

We use the location of the peak in $\langle u_r^* \rangle_s$ as a function of radius to estimate the penetration depth, $\delta^*$, for each model. We define the radius of maximum $\langle u_r^* \rangle_s$ as $r_{\max}$ and calculate the penetration depth as follows

$$\delta^* = r_o - r_{\max}. \tag{3.1}$$

Radial velocity is used to estimate the penetration depth because it has only a single peak that is located centrally within the shear layer, whereas the horizontal components typically have several peaks, with the highest value close to the outer boundary in our $S > 1$ models, see the equatorial sections in figs 3 and 4, and fig. 7 for a representative example of radial velocity profiles. Note that the $\langle \rangle_s$ operation averages any longitudinal dependence of $u_r^*$, as seen in fig. 5 for example. Fig. 8 shows that $\delta^*$ has different behaviour in the two stratification ($S$) regimes, with $\delta^*$ on the order of the shell thickness at low $S$ and obeying power law relationships in $S$ and $B$.

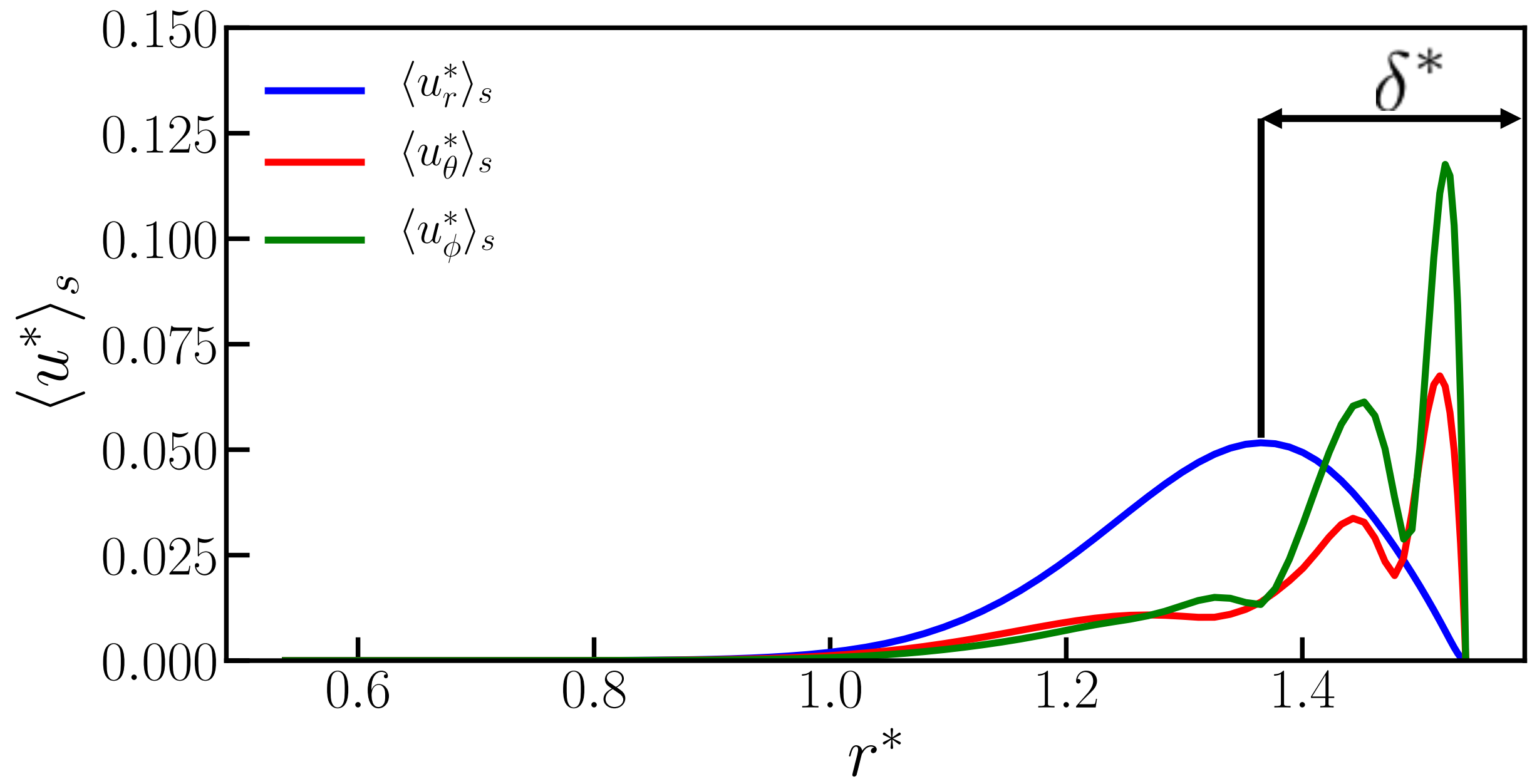

Figure 7: Components of velocity as a function of radius for a model run at $E = 10^{-4}$, $B = 100$ and $S = 1000$. The line colour denotes the flow component (blue for radial, red for meridional and green for azimuthal). The black arrow represents the width used as an estimate for the penetration depth, $\delta^*$, in this model (calculated according to (3.1)).

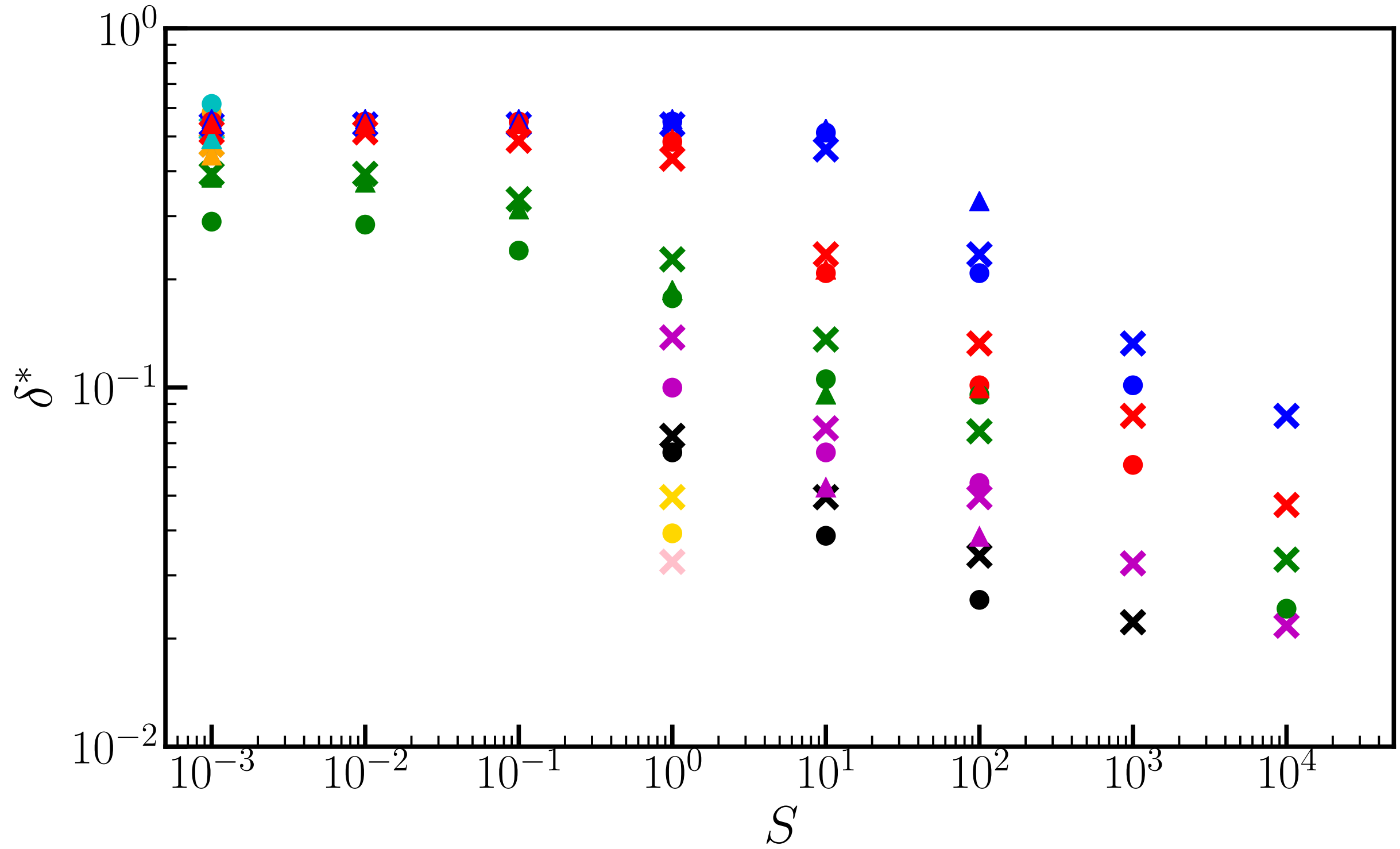

Figure 8: Estimates of the penetration depth $\delta^*$, as a function of the stratification parameter, $S$, for all steady models. Symbol shapes represent the Ekman number, $E$, and colours represent the buoyancy parameter, $B$. The key is given in fig 6a.

## 4. Scaling analysis

In this section, our aim is to recover power laws of the form

$$f = S^a B^b \tag{4.1}$$

from the governing equations to express the velocity components, temperature fluctuations and penetration depth (denoted $f$ above) as functions of the control parameters $S$ and $B$ (and, equivalently, $S$, $Ra_{\mathcal{H}}$ and $E$), where coefficients $a$ and $b$ are to be determined. We then verify these predicted scalings for our models using the volume averaged quantities introduced above, and finally we extrapolate the power laws to planetary core conditions.

### 4.1. *High stratification regime*

At high stratification parameter, $S$, flow is confined to a shear layer of thickness $\delta^*$ at the top of the shell and this penetration depth decreases with increasing stratification. Within the layer, flow tends to be in long, thin lobes with relatively little lateral variation, which suggests that the radial gradients of velocity ($\frac{\partial}{\partial r^*}$) are larger than the horizontal ($\frac{\partial}{\partial \theta}$ and $\frac{\partial}{\partial \phi}$) gradients. Our dimensionless horizontal lengths are $O(1)$ and the relevant radial length scale is $O(\delta^*)$ so that the continuity equation ($\nabla \cdot \boldsymbol{u}$) gives a relationship between the velocity components

$$u_r^* \sim \delta^* u_\theta^* \sim \delta^* u_\phi^*, \tag{4.2}$$

assuming that $\frac{\partial}{\partial \theta} \sim \frac{\partial}{\partial \phi}$. Adherence of our high $S$ models to this scaling was verified using the estimates of $\delta^*$ shown in fig. 8 and volume-averaged velocities $\langle u_r^* \rangle_v$, $\langle u_\theta^* \rangle_v$ and $\langle u_\phi^* \rangle_v$ shown in fig. 6. These results, summarised in fig. B.1 show clear flattening of $\langle u_r^* \rangle_v / \delta^* \langle u_h^* \rangle_v$ for the highest $S$ models, where $\langle u_h^* \rangle_v$ is the average volume-averaged horizontal velocity ($= \frac{1}{2}[\langle u_\theta^* \rangle_v + \langle u_\phi^* \rangle_v]$).

4.1.1. *Vorticity equation balance*

Taking the curl of (2.14) gives the dimensionless vorticity equation for steady flow

$$\frac{\partial \boldsymbol{u^*}}{\partial z^*} = \boldsymbol{\nabla} \times B\ T_1^* \boldsymbol{r^*} + E\nabla^2 \boldsymbol{\omega^*}, \tag{4.3}$$

in which pressure does not appear and inertia is assumed small. In this three-term balance, we note that the buoyancy term is purely horizontal, and so the radial component of the first term must be small except outside the viscous boundary layer. Motivated by the observation that the viscous term is large only near the boundaries (fig. B.2), we seek a thermal wind balance between the horizontal components of the Coriolis and buoyancy terms, and will show subsequently that the resulting scaling remains consistent for cases in which viscosity is also included in the balance. We adopt $\delta^*$ as the relevant length scale in the Coriolis term that controls variations parallel to the rotation axis at high stratification (see fig. 5), and an $O(1)$ horizontal length scale for the buoyancy term since it is determined by the boundary condition. The balance is then

$$\frac{u_{\theta,\phi}^*}{\delta^*} \sim B\ T_1^*. \tag{4.4}$$

The volume-averaged magnitude of the Coriolis and buoyancy terms, scaled by our approximations to those terms using $\delta^*$ and volume-averaged velocities and temperatures ($\frac{\partial \boldsymbol{u^*}}{\partial z^*} \sim \langle u^* \rangle_v / \delta^*$ for Coriolis and $\boldsymbol{\nabla} \times B\ T_1^* \boldsymbol{r^*} \sim B\ \langle T_1^* \rangle_v$ for buoyancy), are plotted for all models in figs B.3 and B.4. These ratios are approximately one for all high $S$ models (excepting a higher value ($\approx 4$) for the model at $S = 100$ and $B = 1$, although we verified this model is converged and otherwise fully consistent with other high $S$ models), and show little $S$ dependence, indicating that the correct scalings are encapsulated in our approximations and that the volume-averaged quantities are suitable diagnostics of model output.

4.1.2. *Temperature equation balance*

The dimensionless time-independent temperature equation is

$$\nabla^2 T_1^* - u_r^* \frac{\partial T_1^*}{\partial r^*} - \frac{u_\theta^*}{r}\frac{\partial T_1^*}{\partial \theta} - \frac{u_\phi^*}{r\sin\theta}\frac{\partial T_1^*}{\partial \phi} - S\ u_r^* r^* = 0. \tag{4.5}$$

Assuming that diffusion occurs on the length scale of the penetration depth, and that the geometric factors of $r$ and $\sin\theta$ are order unity, leaves

$$\frac{T_1^*}{{\delta^*}^2} - 3\frac{u_r^*}{\delta^*} T_1^* - S\ u_r^* \approx 0 \tag{4.6}$$

using the scaling for the velocity components of equation (4.2). For two representative high $S$ models, $(\boldsymbol{u}^* \cdot \nabla)\, T_1^*$ is small compared to the other terms, fig. B.5. Therefore,

$$\frac{T_1^*}{{\delta^*}^2} \sim S\ u_r^*. \tag{4.7}$$

is the appropriate balance. The approximation $\nabla^2 T_1^* \sim \langle T_1^* \rangle_v / {\delta^*}^2$ and the term balance in the temperature equation described by (4.7) were verified for our high $S$ models, see B.6, which shows a clear flattening of $\frac{\langle T_1^* \rangle_v}{{\delta^*}^2} / S\ \langle u_r^* \rangle_v$ for higher stratification parameters and little dependence on $B$.

4.1.3. *Power law scalings*

Rearranging (4.7) for $\delta^*$, eliminating $u_r^*$ using (4.2) from the continuity equation and substituting $B\ T_1^* \delta^*$ for horizontal flow (from balancing $\frac{\partial \boldsymbol{u}^*}{\partial z^*}$ with $\boldsymbol{\nabla} \times B\ T_1^* \boldsymbol{r}^*$ in (4.3)), results in a scaling for the penetration depth in terms of the control parameters

$$\delta^* \sim (S\ B)^{-\frac{1}{4}} \sim (S\ Ra_{\mathcal{H}}\ E)^{-\frac{1}{4}}. \tag{4.8}$$

We now postulate that the radial velocity $u_r^*$ depends on $S$ but not $B$ as it is not directly forced by the thermal wind; it arises to conserve mass for the horizontal velocity components, which are directly forced by the boundary anomalies. Then

$$u_r^* \sim S^a, \tag{4.9}$$

and the horizontal flow components scale as

$$u_{\theta,\phi}^* \sim S^{a+\frac{1}{4}} B^{\frac{1}{4}}, \tag{4.10}$$

from (4.2). The temperature perturbations depend on both $S$ and $B$

$$T_1^* \sim S^b\ B^c \tag{4.11}$$

where the exponents $b$ and $c$ are to be determined. From (4.7) and (4.8),

$$S^{b-a} B^c \sim \frac{T_1^*}{u_r^*} \sim {\delta^*}^2 S \sim S^{\frac{1}{2}} B^{-\frac{1}{2}}, \tag{4.12}$$

substituting (4.9) and (4.11), from which we deduce $c = -\frac{1}{2}$ and $b - a = \frac{1}{2}$. Another assumption is required in order to proceed further with the analysis. We now assume that at sufficiently high $\beta$, the boundary anomalies become unimportant so that the temperature perturbations are independent of $\mathcal{H}$. Then, $T_1^*$ can only depend on the product $S\ B$ and, since the power of $B$ is $-\frac{1}{2}$, the power of $S$ (=$b$) must also be $-\frac{1}{2}$. We have now determined the exponents for the temperature fluctuations

$$T_1^* \sim (SB)^{-\frac{1}{2}} \sim (SRa_{\mathcal{H}}E)^{-\frac{1}{2}}, \tag{4.13}$$

radial flow

$$u_r^* \sim S^{-1}, \tag{4.14}$$

and horizontal flow components

$$u_{\theta,\phi}^* \sim S^{-\frac{3}{4}} B^{\frac{1}{4}} \sim S^{-\frac{3}{4}}\ Ra_{\mathcal{H}}^{\frac{1}{4}}\ E^{\frac{1}{4}}. \tag{4.15}$$

4.1.4. *Empirical fit to models*

In order to test the scaling laws obtained in the previous section, we computed best fits to our models using a least squares inversion of the estimates of the penetration depth and the volume-averaged velocities and temperature perturbations. We seek power laws of the form

$$\tilde{y} = \epsilon S^{\chi} B^{\zeta} \tag{4.16}$$

where the 'observations' $y$ are model outputs, and the predictions $\tilde{y}$ are calculated from the control parameters $S$ and $B$, given the specified functional form. We take the logarithm to transform the power law problem into a linear problem such that

$$\log \tilde{y} = \log \epsilon + \chi \log S + \zeta \log B \tag{4.17}$$

and calculate the prefactor $\epsilon$ and exponents $\chi$ and $\zeta$ using a linear least squares inversion. A summary of the predicted scaling exponents ((4.8) and (4.13)-(4.15)) and those obtained from the least squares fits to all models in the stratification-dominated regime ($S > 1$) is provided in table 1 for comparison. A measure of how well the models are fit is given by the $R^2$ values (rounded to two decimal places throughout). The best fitting exponents are in good agreement with those obtained in the analysis; see also figs 9a to 9c.

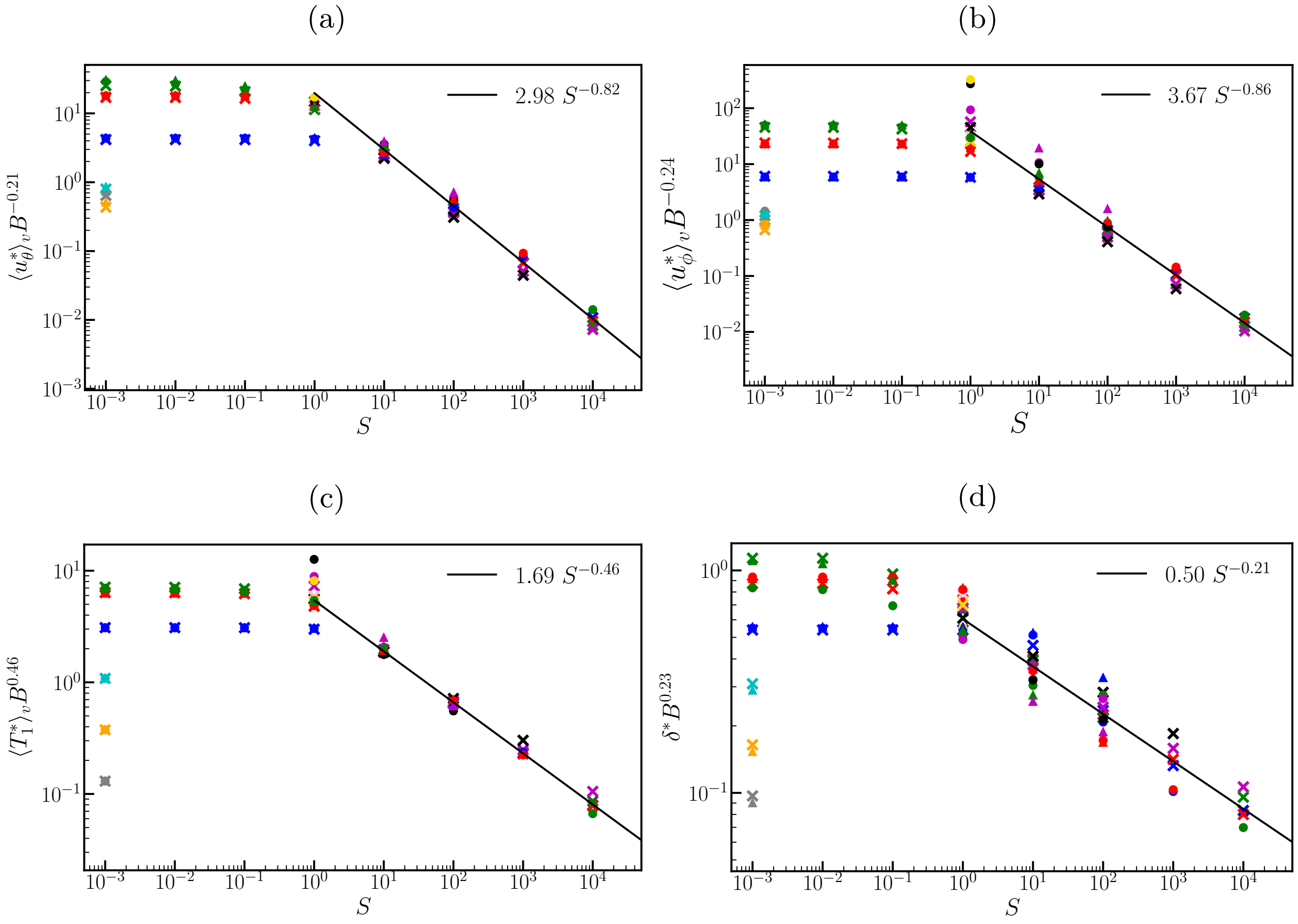

Figure 9: (a) Volume-averaged meridional velocities, (b) volume-averaged azimuthal velocities, (c) volume-averaged temperature perturbations and (d) penetration depth estimates, normalised by the best empirical fit to the buoyancy parameter for all models with $S > 1$, as a function of $S$. Symbol shapes represent the Ekman number, $E$, and colours represent the buoyancy parameter, $B$. The key is given in fig 6a. The black line shows the best fitting power law for $S$ in the stratification-dominated regime.

| Quantity | Prediction | Fit to models | Fit $R^2$ |
|---|---|---|---|
| $u_r^*$ | $S^{-1}$ | $S^{-1.01}B^{0.01}$ | 0.98 |
| $u_\phi^*$ | $S^{-\frac{3}{4}}B^{\frac{1}{4}}$ | $S^{-0.86}B^{0.24}$ | 0.97 |
| $u_\theta^*$ | $S^{-\frac{3}{4}}B^{\frac{1}{4}}$ | $S^{-0.82}B^{0.21}$ | 0.99 |
| $T_1^*$ | $S^{-\frac{1}{2}}B^{-\frac{1}{2}}$ | $S^{-0.46}B^{-0.46}$ | 1.00 |
| $\delta^*$ | $S^{-\frac{1}{4}}B^{-\frac{1}{4}}$ | $S^{-0.21}B^{-0.23}$ | 0.95 |

Table 1: Scaling analysis and least squares inversion results for all $S > 1$ models.

#### 4.1.5. *The role of viscosity*

Having verified our two-term balance in the vorticity equation, we now address the question of whether our scalings are also consistent when considering all three terms.

The additional viscous term scales as $E\ u^*\ l_\nu^{-3}$, where $l_\nu$ is a relevant length scale yet to be determined.

The assertion that $l_\nu = \delta$ leads to $l_\nu = \delta \sim E^{1/2}$ independent of $S$, which as figure 7 demonstrates is not the case as $\delta$ has clear empirical $S$-dependence (see also fig. B.7, which shows the ratio of the viscous term to the incorrect scaling $E\ u_h^* {\delta^*}^{-3}$ as a function of $S$ for all models). Alternatively, assuming that $l_\nu$ represents a thin boundary layer (consistent with figure B.2), then the three-term balance determines $l_\nu$ to be

$$l_\nu \sim (E\ \delta^*)^{\frac{1}{3}}. \tag{4.18}$$

Fig. 10 shows that the shear layer thickness (given in (4.8)) and the Ekman layer depth are comparable for most of our models, which are therefore are in fact described by a three-term (rather than a two-term) balance within the shear layer. The inclusion of viscosity within the balance in no way invalidates our analysis of the two-term scaling, but merely provides information about the characteristic lengthscale $l_\nu$ at which viscosity becomes important. Indeed, our derived scalings of the previous sections, confirmed empirically, appear to hold independently of the relative size of $l_\nu$ and $\delta^*$. It is worth pointing out the physically relevant planetary regime is one in which $E \ll 1$ and $l_\nu \ll \delta$ (see also section 5), and therefore in this limit the two term balance is appropriate for the shear-layer. We speculate that there may be a different behavioural regime in which $l_\nu \gg \delta^*$ for certain choices of parameters, when viscosity balances just one other term.

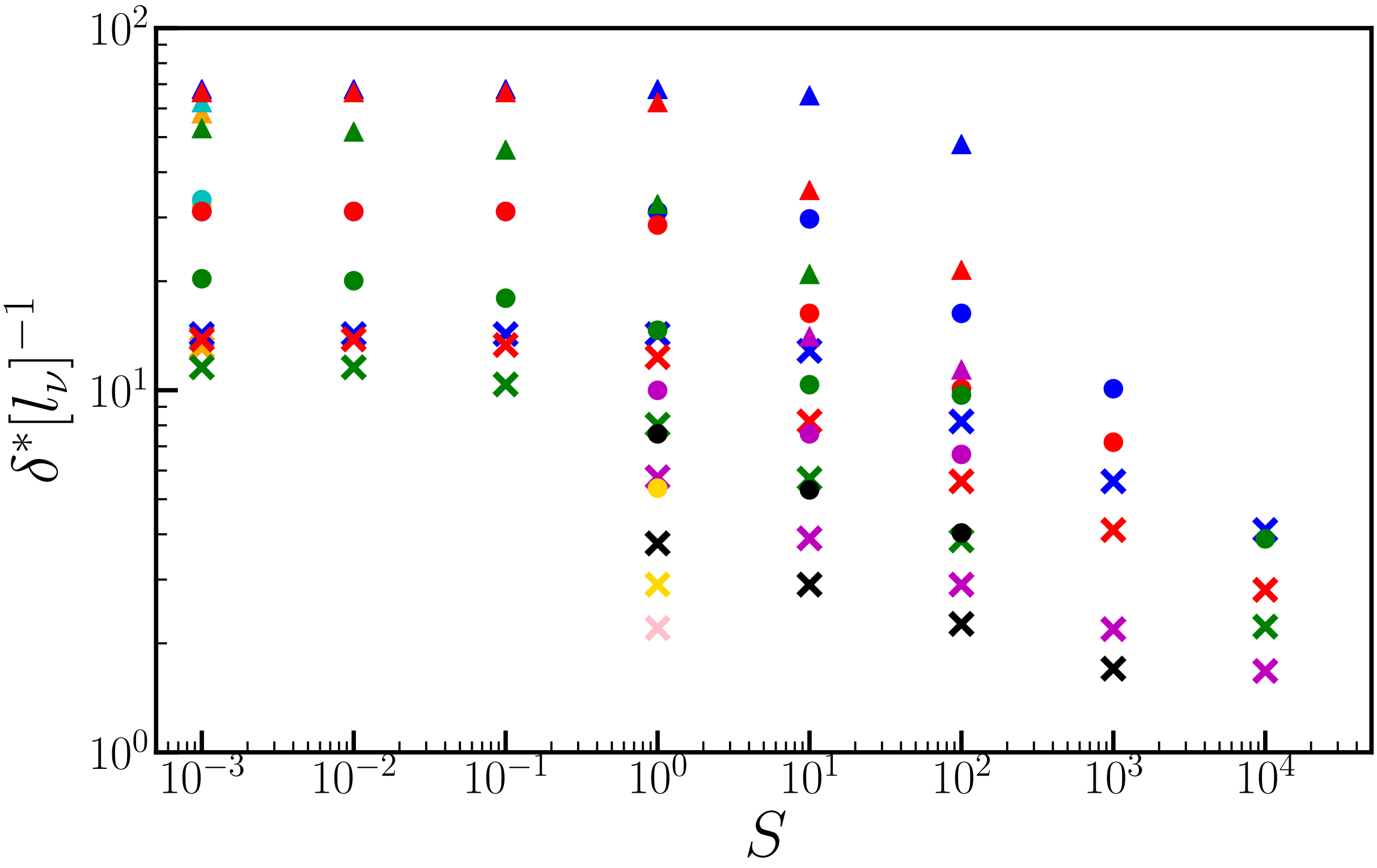

Figure 10: The shear layer thickness, $\delta^*$, scaled by the dimensionless Ekman layer thickness estimated using (4.18) as a function of the stratification parameter, $S$, for all steady models. Symbol shapes represent the Ekman number, $E$, and colours represent the buoyancy parameter, $B$. The key is given in fig 6a.

### 4.2. *Low stratification regime*

At low stratification, $S$, the velocities and temperature perturbations do not depend on $S$, and flow occupies the whole shell rather than being concentrated to a thin layer (i.e. $\delta^* \sim O(1)$). The dynamics of this regime have previously been investigated in Zhang & Gubbins (1992), Gibbons & Gubbins (2000) and Gibbons *et al.* (2007), but are further explored here using a much broader $(E, S, B)$ parameter space than prior works.

As in the high $S$ regime, the largest terms in the thermal wind balance are Coriolis and buoyancy, with viscous effects only important in the mechanical boundary layers; see fig B.8. We again consider the horizontal component of $\frac{\partial \boldsymbol{u}^*}{\partial z^*} \sim \boldsymbol{\nabla} \times B\ T_1^* \boldsymbol{r}^*$ from (4.3). We use the large $O(1)$ length scale in the Coriolis term due to the lack of variation parallel to the rotation axis (shown in the meridional sections of fig 5) and in the buoyancy term so that the latter term is again approximated as $\boldsymbol{\nabla} \times B\ T_1^* \boldsymbol{r^*} \sim B\ \langle T_1^* \rangle_v$. This leaves

$$u_r^* \sim u_h^* \sim B\ T_1^*, \tag{4.19}$$

since the radial and horizontal velocity components are assumed to follow the same scalings in this regime ($u_r^* \sim u_h^*$).

In the low $S$ regime, the dominant terms of the temperature equation are those for diffusion and advection of the temperature perturbations so that

$$\nabla^2 T_1^* \sim (\boldsymbol{u}^* \cdot \nabla)\, T_1^*, \tag{4.20}$$

see fig B.9. The term $(\boldsymbol{u}^* \cdot \nabla)\, T_1^*$ is problematic to estimate as $\langle u^* T_1^* \rangle_v \neq \langle u^* \rangle_v \langle T_1^* \rangle_v$. In any case, when we postulate that the velocities and temperature perturbations depend only on $B$ in the low $S$ regime such that $u^* \sim B^a$ and $T_1^* \sim B^b$, where exponents $a$ and $b$ are to be determined, equations (4.19) and (4.20) alone do not provide enough information to determine the $B$ exponents explicitly. Indeed, it is not even clear whether scaling laws comparable to those for the high $S$ case exist, since high $B$ values in the low $S$ regime encompass a mixture of steady and unsteady solutions (see fig. 1), which cannot necessarily be expected to scale in the same way. Therefore, rather than formulating the scaling laws in terms of the buoyancy parameter, we use the mean kinetic energy equation to recover scalings in terms of the buoyant power in the next section.

#### 4.2.1. *Scaling laws*

The mean kinetic energy equation, for no-slip boundary conditions as applied in our simulations, is obtained by taking the scalar product of the velocity $\boldsymbol{u^*}$ with the momentum equation (2.14), averaging over time (denoted with overlines) and integrating over the fluid shell volume $V$ (e.g. King & Buffett 2013)

$$\int_V B\, \overline{T_1^* u_r^*}\, dV - \int_V E\, \overline{\omega^*}^2\, dV = 0. \tag{4.21}$$

The buoyant power $P = \int_V B\, \overline{T_1^* u_r^*}\, dV$ is expended by the viscous dissipation $D_\nu = \int_V E\, \overline{\omega^*}^2\, dV$. Assuming that the vorticity scales as $u^*/l_\nu$, where $l_\nu$ is given by (4.18), equation (4.21) gives a scaling relation for the characteristic velocity in terms of the buoyant power. Note that the Reynolds number $Re$ is equal to the thermal Péclet number $Pe$ since $Pr = 1$ in all models so that it represents a characteristic velocity of the final steady-state $Re = Pe = u^*$ so that

$$Re \sim P^{\frac{1}{2}}. \tag{4.22}$$

In the absence of a magnetic field (and therefore Ohmic dissipation), equation (4.21) should hold for all simulations in this work regardless of parameters. Fig. 11 and the least-squares fit to the models ($Re \sim P^{0.50}$ with $R^2 = 0.99$) shows that this is indeed the case.

### 4.3. *Effects of the shell aspect ratio*

We have used an aspect ratio $\eta = 0.35$ in all previous models, however as we would like to apply the derived scaling laws to other shells with different aspect ratios, we now consider whether varying the geometry influences the results. To this end, we have run simulations with $\eta = 0.01$ using the parameters listed in table A.4 and obtained steady-state solutions. It is apparent that the overall dynamics of the low aspect ratio models is very similar to the previously presented models, fig. C.1. We again have two stratification regimes, a low $S$ regime in which dynamics occupy the entire shell and buoyancy is the dominant effect, and a high $S$ regime in which stratification dominates and dynamics are concentrated towards the outer boundary. In both regimes, the phase of the velocity and temperature lobes with respect to the boundary anomaly pattern is the same as in the previously discussed models. We have computed the best empirical fits to the high $S$ models in this geometry (shown in fig. 12) and confirm that these models obey the same scaling laws as derived in 4.1.3. Note that the values of the quantities shown in figures C.1 and 12, are different from those shown in previous sections for the same apparent parameter values because the length scales in the parameters $S$ and $B$ differ because $d = r_o - r_i = r_o(1 - \eta)$, and averaging takes place over different volumes, meaning that for example, $B = 1$ and $St = 1000$ models at $\eta = 0.35$ and $\eta = 0.01$ are not directly comparable without accounting for geometric factors. It is worth remarking that the theoretical scaling for the horizontal velocity components (which scale as $\sim S^{-3/4}$) actually agree slightly better with the numerics in the quasi-full sphere than the spherical shell calculations, indicating a possible weak dependence on $r_i$ for such quantities.

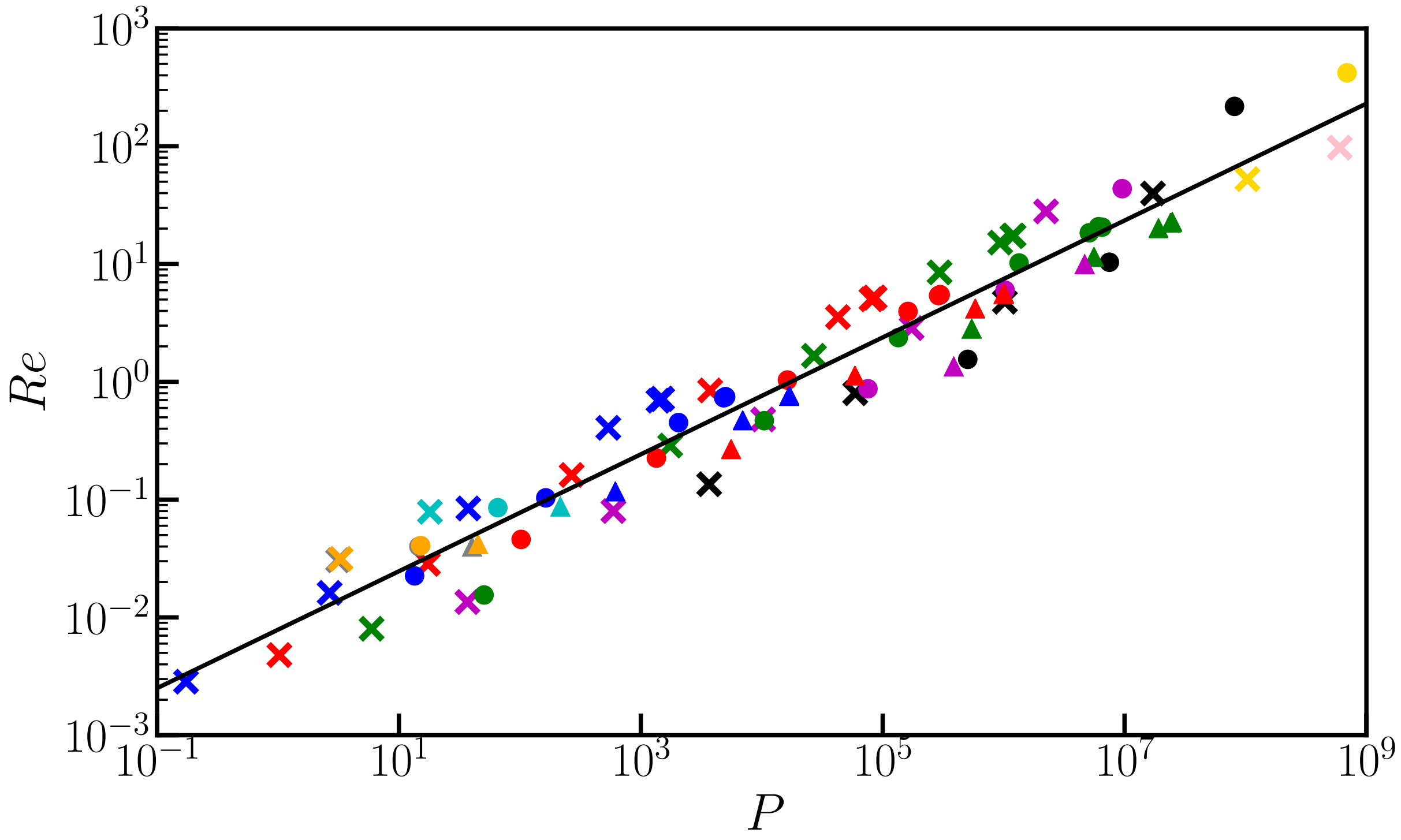

Figure 11: The Reynolds number $Re$ as a function of the buoyant power $P$ for all steady models. The black line shows the best empirical fit to the models, which is $Re \sim P^{0.50}$ with $R^2 = 0.99$. Symbol shapes represent the Ekman number, $E$, and colours represent the buoyancy parameter, $B$. The key is given in fig 6a.

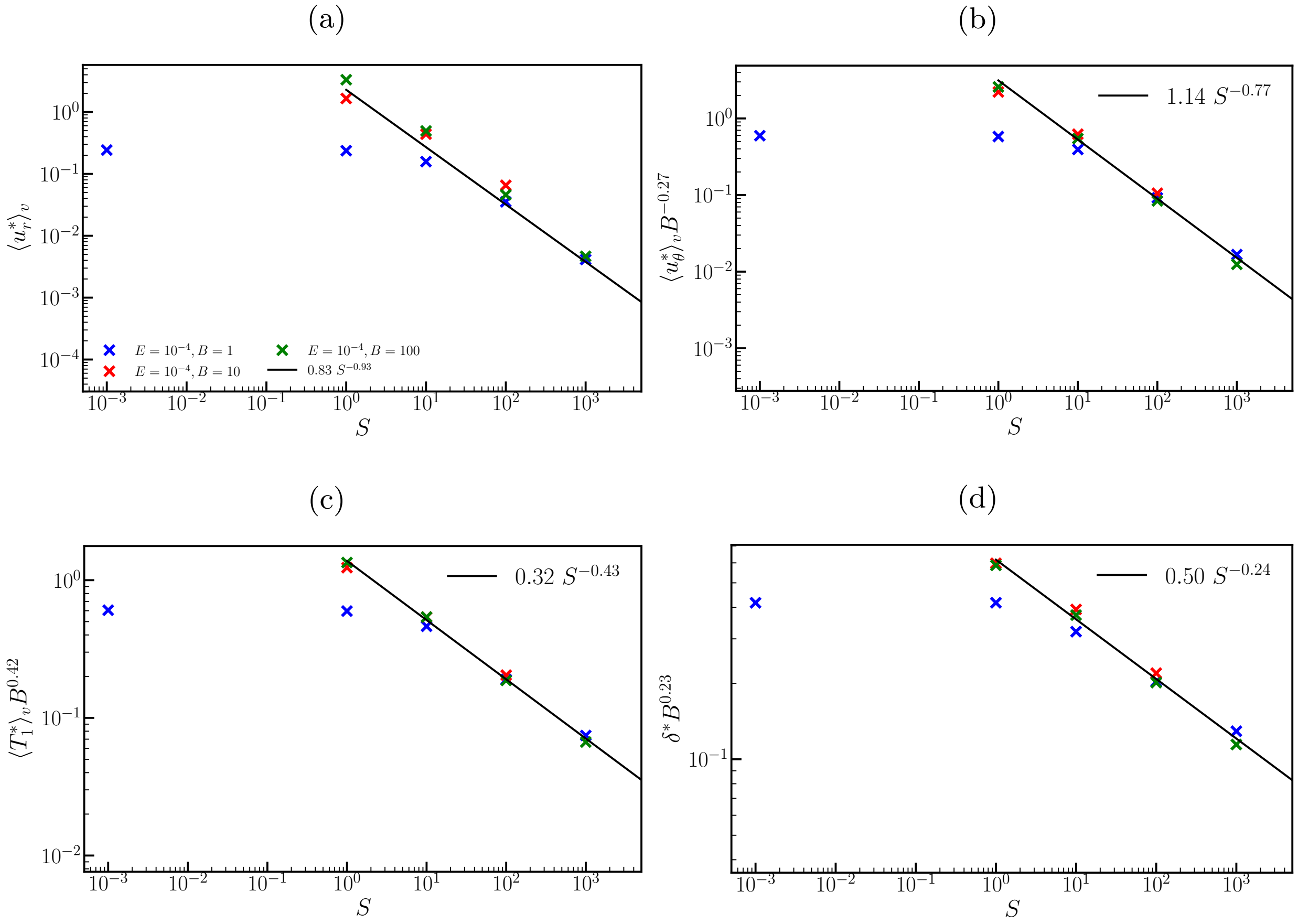

Figure 12: (a) Volume-averaged radial velocities, (b) volume-averaged azimuthal velocities, (c) volume-averaged temperature perturbations and (d) penetration depth estimates, normalised by the best empirical fit to the buoyancy parameter for all models with $\eta = 0.01$ and $S > 1$, as a function of $S$. The $R^2$ values for the fits are, respectively, 0.95, 0.99, 1.00 and 0.99. Symbol shapes represent the Ekman number, $E$, and colours represent the buoyancy parameter, $B$. The black line shows the best fitting power law in $S$ for models at $S > 1$.

## 5. Application of scaling laws to planetary cores

In order to apply our power law scalings to a planet, we must estimate $S$ and $B$ for its outer core. We write $\beta$ and $\mathcal{H}$ in terms of temperature gradients at the CMB

$$\beta\, d = \left.\frac{\mathrm{d}T_{ad}}{\mathrm{d}r}\right|_{r_o} - \left.\frac{\mathrm{d}T_c}{\mathrm{d}r}\right|_{r_o} \tag{5.1}$$

where $T_{ad}$ is the adiabatic temperature and $T_c$ is the core temperature at the CMB, and

$$\mathcal{H} = \left.\frac{\mathrm{d}T'}{\mathrm{d}r}\right|_{r_o} = \frac{q'}{k_m}, \tag{5.2}$$

where $T'$ ($q'$) is the anomalous temperature (heat flow per unit area) on the CMB and $k_m$ is the lower mantle thermal conductivity. Note that $\mathcal{H}$ is related to the mantle-side temperature variations, and not the core-side temperature variations, because the mantle imposes the CMB heat flux on the core. The gradients in (5.1) are evaluated using

$$\left.\frac{\mathrm{d}T_{ad}}{\mathrm{d}r}\right|_{r_o} = \frac{\alpha_T g_c T_c}{C_p} \tag{5.3}$$

and

$$\left.\frac{\mathrm{d}T_c}{\mathrm{d}r}\right|_{r_o} = \frac{Q_{cmb}}{A_{cmb} k_c} \tag{5.4}$$

where $g_c$ is the acceleration due to gravity at the CMB, $C_p$ is the core specific heat, $Q_{cmb}$ is the total CMB heat flux, $A_{cmb}$ is the area of the CMB (=$4\pi r_o^2$) and $k_c$ is the core thermal conductivity. For the Earth's core, we have taken a range of plausible values from the literature, given in table 2, and calculated a range of possible $S$ and $B$ parameters.

Estimating the stratification parameter ($S = \beta d/\mathcal{H}$) is particularly challenging due to large uncertainties on $\mathcal{H}$, the magnitude of lateral variations in CMB heat flux, whose estimate derives from relating observed shear-wave anomalies with either thermal or chemical heterogeneities. If the anomalies are attributed predominantly to thermal differences in the mantle, then the value of $q'$ from table 2 leads to $S$ values about $O(10^{-7})$ to $O(10^{-5})$ and $B$ values of $O(10^{18})$ to $O(10^{19})$ placing the core in a regime in which the stratified layer is likely penetrated by unsteady boundary-driven flow.

On the other hand, if the mantle heterogeneities are attributed instead to chemical anomalies (e.g. Garnero *et al.* 2016; Lau *et al.* 2017), then $\mathcal{H}$ could be much smaller than the above estimate, rendering $S$ plausibly $O(1)$ or above, placing the core in the stratification-dominated regime. Taking $S = 1$ for illustration with the above estimates of $B$, applying the high $S$ scalings (4.13) to (4.15) gives dimensional temperature perturbations of $10^{-3}\,\mathrm{K}$ to $10^{-2}\,\mathrm{K}$, radial velocities of $10^{-12}\,\mathrm{m\,s^{-1}}$, horizontal velocities of $10^{-7}\,\mathrm{m\,s^{-1}}$ and penetration depths of 10 m to 40 m, much thicker than the estimated viscosity boundary layer in Earth's core of about 1 m (e.g. Livermore *et al.* 2016) associated with $E = 10^{-15}$. A similar analysis for Ganymede's core, using values from table 1 of Rückriemen *et al.* (2015) and estimating $\alpha_T = 5.8 \times 10^{-5}$ based on Williams & Nimmo (2004), gives $B \sim O(10^{12} - 10^{13})$ and $S \sim O(10^{-2} - 10^{-1})$, assuming the mantle heterogeneities are attributed thermal anomalies. As for the Earth, if the anomalies are predominantly due to chemical sources, these $S$ values are significantly underestimated and Ganymede's core will be in the stratified regime.

For comparison with other works on stratified fluids, it is of interest to calculate the Brunt-Väisälä frequency, $N$, defined by

$$N^2 = -\frac{g}{\rho_0}\frac{\partial \rho'}{\partial r} \tag{5.5}$$

both for our models and for the planetary interiors considered. Non-dimensionalising with the same scalings as used previously gives the ratio of the Brunt-Väisälä frequency to the rotation rate

$$\frac{N}{2\Omega} = \sqrt{\frac{B\ E}{Pr}\frac{\partial T^*}{\partial r^*}} = \sqrt{\frac{B\ E\ S}{Pr}}, \tag{5.6}$$

assuming $\frac{\partial T^*}{\partial r^*} \approx \frac{\partial T_0^*}{\partial r^*}$ due to the small magnitudes of the temperature perturbations.

Values of this ratio for our simulations vary between $\mathrm{O}(10^{-6})$ and $\mathrm{O}(10)$, given in tables A.1 to A.4 in appendix A. Based on our $B - S$ estimates for Earth and Ganymede, along with $E$ and $Pr$ estimates from table 4 of Schubert & Soderlund (2011), we estimate their Brunt-Väisälä ratios both of $O(1)$ for some parameter combinations, consistent with other estimates using different methods (e.g. Buffett 2014). Ignoring the dependence on $E$, it is worth remarking that the relationship between $N$ and the product $S\,B$ may explain why this quantity is so important in our derived theoretical scalings, with $\delta^* \sim N^{-1/4}$ and $T_1^* \sim N^{-1/2}$.

| Parameter | Symbol | Value | Reference |
|---|---|---|---|
| Inner core radius | $r_i$ | $1221\,\mathrm{km}$ | Dziewonski & Anderson (1981) |
| Outer core radius | $r_o$ | $3480\,\mathrm{km}$ | Dziewonski & Anderson (1981) |
| Shell thickness | $d\ (= r_o - r_i)$ | $2259\,\mathrm{km}$ | Dziewonski & Anderson (1981) |
| Gravitational acceleration constant at CMB | $g_c$ | $10.68\,\mathrm{ms}^{-2}$ | Olson (2009) |
| Angular velocity of rotation | $\Omega$ | $7.272 \times 10^{-5}\,\mathrm{s}^{-1}$ | Olson (2009) |
| Coefficient of thermal expansion | $\alpha_T$ | $1.5 \times 10^{-5}\,\mathrm{K}^{-1}$ | Gubbins *et al.* (2003) |
| Core thermal diffusivity | $\kappa$ | $1.25 \times 10^{-5}\,\mathrm{m}^2\,\mathrm{s}^{-1}$ | Pozzo *et al.* (2012) |
| Core thermal conductivity | $k_c$ | $100\,\mathrm{W\,m}^{-1}\,\mathrm{K}^{-1}$ | Pozzo *et al.* (2013) |
| Lower mantle thermal conductivity | $k_m$ | $10\,\mathrm{W\,m}^{-1}\,\mathrm{K}^{-1}$ | Ammann *et al.* (2014) |
| Core specific heat capacity | $C_p$ | $728\,\mathrm{J\,kg}^{-1}\,\mathrm{K}^{-1}$ | Gubbins *et al.* (2003) |
| CMB temperature | $T_c$ | $4000\,\mathrm{K}$ | Olson (2009) |
| Total CMB heat flow | $Q_{cmb}$ | $5\,\mathrm{TW}$ to $17\,\mathrm{TW}$ | Lay *et al.* (2008); Nimmo (2015) |
| Total adiabatic heat flow | $Q_{ad}$ | $14\,\mathrm{TW}$ to $16\,\mathrm{TW}$ | Pozzo *et al.* (2012) |
| Peak-to-peak anomalous CMB heat flow | $q'$ | $100\,\mathrm{mWm}^{-2}$ to $500\,\mathrm{mWm}^{-2}$ | Nakagawa & Tackley (2013) |

Table 2: Outer core and lower mantle physical, thermodynamics and transport properties used to estimate $S$ and $B$ for the Earth.

## 6. Discussion and conclusions

We have investigated a thermally stratified fluid in a rotating spherical shell subject to a laterally varying heat flux pattern on the outer boundary. Converged, steady-state numerical simulations were obtained for $Pr = 1$, $E = 10^{-6}$ to $E = 10^{-3}$, $S = 10^{-3}$ to $S = 10^{4}$ and $B = 10^{-3}$ to $B = 10^{6}$. For some parameters, we obtained time-dependent solutions, which were not analysed in this study, however we were able to map the stability domain in parameter space in greater detail than any previous study. The steady-state solutions separate into two distinct dynamical regimes corresponding to low stratification parameter ($S$), in which buoyancy effects dominate the dynamics, and high $S$, in which stratification effects dominate.

In the low $S$ regime, the inhomogeneous thermal boundary condition drives flows that are locked to the boundary pattern and penetrate most of the shell thickness. We have determined a power law dependency of the characteristic velocity $Re$ as a function of the buoyant power. In the high $S$ regime, stratification strongly suppresses radial flow but horizontal flow is less affected. All flow is concentrated toward the outer boundary, resulting in shear layers whose thickness decreases with increasing $B$ and $S$. This layer thickness represents the depth to which the boundary driven flows penetrate the stratified fluid. We have developed scaling relations for the velocity components, temperature perturbations and penetration depth as functions of the control parameters $E$, $B$ and $S$; these are summarised in table 1.

We have used these scaling relationships to extrapolate to Earth's core using a range of plausible parameters. If the Earth's mantle heterogeneities are attributed to thermal anomalies, the outer core is in the buoyancy-dominated regime and no steady-state solutions exist. In that case, it is likely that unsteady boundary-driven flows can penetrate the stratified layer. On the other land, if such heterogeneities are linked to chemical anomalies (e.g. Garnero *et al.* 2016; Lau *et al.* 2017), the much reduced heat-flux boundary condition would likely place Earth's core in the stratification-dominated regime where penetration from steady boundary-driven flows is not possible. In that case, the shear layer thickness (i.e. the depth of penetration of boundary driven flows through the core) is very small (on the order of a few tens of metres) compared to the stable layer thickness and the predicted velocities are several orders of magnitude smaller than those inferred from inversions of geomagnetic secular variation (e.g. Holme 2015). Since there is no reason why the 'observed' flows have to be generated (even in part) by mantle heterogeneities, the high $S$ scalings suggest that we observe general convective flow rather than boundary-driven flow. Furthermore, it seems unlikely that chemical anomalies in the lowermost mantle are able to directly affect the magnetic field that is generated inside the core (by creating persistent non-zonal features for example) through steady boundary-driven flows.

However, the relative contributions of thermal and chemical anomalies to the boundary forcing is poorly constrained for Earth and not at all for other bodies (including Ganymede), hence the difficulty in estimating $\mathcal{H}$ and the resulting uncertainty as to which stratification regime their outer cores belong. Interestingly, this means that independent evidence of penetrating flow within the stable layer, for example through the magnetic signature of upwellings and patches of reversed magnetic flux (Gubbins 2007; Metman *et al.* 2018), may be able to discriminate between these two regimes and therefore offer evidence that constrains the heat-flux on the boundary, and therefore mantle composition.

Finally, we have considered steady-state solutions in entirely stratified spherical shells with no convection or magnetic field generation; further work is needed to investigate the effects of adding these dynamics to our simplified models. The fluid dynamics problem studied here should be relevant in the uppermost region of the outer core, where no convection is expected due to stratification. Yet, it is possible that at sufficiently high $B$, models at $S = 1$ (the lowest stratification parameter required for our high $S$ scalings to be applicable, and a plausible value for Earth's outer core) will be unsteady rather than steady. This transition may well occur at a $B$ lower than our estimates for Earth's core, however, computational limitations have prevented us from reaching this transition and our simulations remain many orders of magnitude from Earth estimates. Since our systematic parameter study has revealed the different dynamical regimes that exist in the absence of internal convection, future studies will be able to benchmark against the present results and also target particular regions of parameter space to make most effective use of available computational resources.

We thank three anonymous referees for comments and suggestions that helped to improve the manuscript. This study was initiated by J.S. as his final year undergraduate project. G.A.C. was supported by NERC grant NE/M012190/1, C.J.D. is supported by a NERC Independent Research Fellowship (NE/L011328/1) and P.W.L. was partially supported by NERC grant NE/G014043/1. This work was undertaken on ARC2, part of the High Performance Computing facilities at the University of Leeds, UK. Contour plots were produced in VisIt (Childs *et al.* 2012), using scripts written by Jon Mound. Other figures were produced using Python's Matplotlib (Hunter 2007).

## Appendix A. Summary tables

Summary tables of the model resolution, control parameters and selected output parameters for all simulations. In all cases $Pr = 1$ and the shell aspect ratio $\eta = 0.35$ for models in tables A.1 to A.3 and $\eta = 0.01$ for models in table A.4. Definitions for $B$, $S$ and $Ra_{\mathcal{H}}$ are given in 2.1. The quantity $N/2\Omega$, defined in (5.6), is the ratio of the Brunt-Väisälä frequency, $N$, to the rotation rate $\Omega$. The variable $n_r$ is the number of radial points within the fluid shell, $l_{max}$ is the maximum degree of the spherical harmonic expansion ($=m_{max}$, the maximum order of the expansion). Since $Re = Pe = \langle u^* \rangle_v$, the Rossby number is

$$Ro = 2\ Re\ E = 2\langle u^* \rangle_v\ E. \quad \text{(A 1)}$$

| $B$ | $S$ | $Ra_{\mathcal{H}}$ | $n_r$ | $l_{max}$ | $\frac{N}{2\Omega}$ | $Re$ | $Ro$ | State |
|---|---|---|---|---|---|---|---|---|
| 0.001 | 0.001 | 10 | 32 | 32 | $1.00\times10^{-5}$ | 0.03 | $6.11\times10^{-6}$ | steady |
| 0.01 | 0.001 | $10^2$ | 32 | 32 | $3.16\times10^{-5}$ | 0.03 | $6.28\times10^{-6}$ | steady |
| 0.1 | 0.001 | $10^3$ | 32 | 32 | $1.00\times10^{-4}$ | 0.08 | $1.58\times10^{-5}$ | steady |
| 1 | 0.001 | $10^4$ | 60 | 48 | $3.16\times10^{-4}$ | 0.72 | $1.44\times10^{-4}$ | steady |
| 1 | 0.01 | $10^4$ | 60 | 48 | $1.00\times10^{-3}$ | 0.72 | $1.44\times10^{-4}$ | steady |
| 1 | 0.1 | $10^4$ | 60 | 48 | $3.16\times10^{-3}$ | 0.72 | $1.43\times10^{-4}$ | steady |
| 1 | 1 | $10^4$ | 60 | 48 | $1.00\times10^{-2}$ | 0.69 | $1.38\times10^{-4}$ | steady |
| 1 | 10 | $10^4$ | 60 | 48 | $3.16\times10^{-2}$ | 0.41 | $8.14\times10^{-5}$ | steady |
| 1 | 100 | $10^4$ | 60 | 48 | $1.00\times10^{-1}$ | 0.08 | $1.68\times10^{-5}$ | steady |
| 1 | 1000 | $10^4$ | 60 | 48 | $3.16\times10^{-1}$ | 0.02 | $3.23\times10^{-6}$ | steady |
| 1 | 10000 | $10^4$ | 60 | 48 | 1.00 | 0.003 | $5.69\times10^{-7}$ | steady |
| 10 | 0.001 | $10^5$ | 60 | 48 | $1.00\times10^{-3}$ | 5.19 | $1.04\times10^{-3}$ | steady |
| 10 | 0.01 | $10^5$ | 60 | 48 | $3.16\times10^{-3}$ | 5.18 | $1.03\times10^{-3}$ | steady |
| 10 | 0.1 | $10^5$ | 60 | 48 | $1.00\times10^{-2}$ | 5.02 | $1.00\times10^{-3}$ | steady |
| 10 | 1 | $10^5$ | 60 | 48 | $3.16\times10^{-2}$ | 3.55 | $7.10\times10^{-4}$ | steady |
| 10 | 10 | $10^5$ | 60 | 48 | $1.00\times10^{-1}$ | 0.84 | $1.69\times10^{-4}$ | steady |
| 10 | 100 | $10^5$ | 60 | 48 | $3.16\times10^{-1}$ | 0.16 | $3.24\times10^{-5}$ | steady |
| 10 | 1000 | $10^5$ | 60 | 48 | 1.00 | 0.03 | $5.70\times10^{-6}$ | steady |
| 10 | 10000 | $10^5$ | 80 | 64 | 3.16 | 0.005 | $9.58\times10^{-7}$ | steady |
| 100 | 0.001 | $10^6$ | 96 | 96 | $3.16\times10^{-3}$ | 17.45 | $3.49\times10^{-3}$ | steady |
| 100 | 0.01 | $10^6$ | 96 | 96 | $1.00\times10^{-2}$ | 17.25 | $3.45\times10^{-3}$ | steady |
| 100 | 0.1 | $10^6$ | 80 | 64 | $3.16\times10^{-2}$ | 15.21 | $3.04\times10^{-3}$ | steady |
| 100 | 1 | $10^6$ | 80 | 64 | $1.00\times10^{-1}$ | 8.50 | $1.70\times10^{-3}$ | steady |
| 100 | 10 | $10^6$ | 80 | 64 | $3.16\times10^{-1}$ | 1.66 | $3.33\times10^{-4}$ | steady |
| 100 | 100 | $10^6$ | 80 | 64 | 1.00 | 0.29 | $5.76\times10^{-5}$ | steady |
| 100 | 10000 | $10^6$ | 224 | 224 | 10.0 | 0.008 | $1.60\times10^{-6}$ | steady |
| 1000 | 0.001 | $10^7$ | 256 | 256 | | | | unsteady |
| 1000 | 0.01 | $10^7$ | 96 | 96 | | | | unsteady |
| 1000 | 0.1 | $10^7$ | 160 | 160 | | | | periodic |
| 1000 | 1 | $10^7$ | 96 | 96 | $3.16\times10^{-1}$ | 27.99 | $5.60\times10^{-3}$ | steady |
| 1000 | 10 | $10^7$ | 96 | 96 | 1.00 | 2.86 | $5.72\times10^{-4}$ | steady |
| 1000 | 100 | $10^7$ | 64 | 64 | 3.16 | 0.48 | $9.59\times10^{-5}$ | steady |
| 1000 | 1000 | $10^7$ | 192 | 192 | 10.0 | 0.08 | $1.60\times10^{-5}$ | steady |
| 1000 | 10000 | $10^7$ | 224 | 224 | 31.6 | 0.01 | $2.70\times10^{-6}$ | steady |
| 10000 | 1 | $10^8$ | 64 | 64 | 1.00 | 39.73 | $7.95\times10^{-3}$ | steady |
| 10000 | 10 | $10^8$ | 64 | 64 | 3.16 | 4.79 | $9.58\times10^{-4}$ | steady |
| 10000 | 100 | $10^8$ | 128 | 128 | 10.0 | 0.80 | $1.60\times10^{-4}$ | steady |
| 10000 | 1000 | $10^8$ | 64 | 64 | 31.6 | 0.14 | $2.70\times10^{-5}$ | steady |
| 100000 | 1 | $10^9$ | 64 | 64 | 3.16 | 52.57 | $1.05\times10^{-2}$ | steady |
| 1000000 | 1 | $10^{10}$ | 96 | 96 | 10.0 | 97.37 | $1.95\times10^{-2}$ | steady |

Table A.1: Summary of all numerical simulations with $E = 10^{-4}$.

| $B$ | $S$ | $Ra_{\mathcal{H}}$ | $n_r$ | $l_{max}$ | $\frac{N}{2\Omega}$ | $Re$ | $Ro$ | State |
|---|---|---|---|---|---|---|---|---|
| 0.001 | 0.001 | $10^2$ | 48 | 48 | $3.16\times10^{-6}$ | 0.04 | $8.01\times10^{-7}$ | steady |
| 0.01 | 0.001 | $10^3$ | 48 | 48 | $1.00\times10^{-5}$ | 0.04 | $8.15\times10^{-7}$ | steady |
| 0.1 | 0.001 | $10^4$ | 48 | 48 | $3.16\times10^{-5}$ | 0.09 | $1.71\times10^{-6}$ | steady |
| 1 | 0.001 | $10^5$ | 48 | 48 | $1.00\times10^{-4}$ | 0.75 | $1.49\times10^{-5}$ | steady |
| 1 | 0.01 | $10^5$ | 48 | 48 | $3.16\times10^{-4}$ | 0.75 | $1.49\times10^{-5}$ | steady |
| 1 | 0.1 | $10^5$ | 48 | 48 | $1.00\times10^{-3}$ | 0.75 | $1.49\times10^{-5}$ | steady |
| 1 | 1 | $10^5$ | 48 | 48 | $3.16\times10^{-3}$ | 0.73 | $1.46\times10^{-5}$ | steady |
| 1 | 10 | $10^5$ | 64 | 64 | $1.00\times10^{-2}$ | 0.45 | $9.02\times10^{-6}$ | steady |
| 1 | 100 | $10^5$ | 64 | 64 | $3.16\times10^{-2}$ | 0.10 | $2.07\times10^{-6}$ | steady |
| 1 | 1000 | $10^5$ | 64 | 64 | 0.1 | 0.02 | $4.51\times10^{-7}$ | steady |
| 10 | 0.001 | $10^6$ | 48 | 48 | $3.16\times10^{-4}$ | 5.49 | $1.10\times10^{-4}$ | steady |
| 10 | 0.01 | $10^6$ | 48 | 48 | $1.00\times10^{-3}$ | 5.48 | $1.10\times10^{-4}$ | steady |
| 10 | 0.1 | $10^6$ | 48 | 48 | $3.16\times10^{-3}$ | 5.40 | $1.08\times10^{-4}$ | steady |
| 10 | 1 | $10^6$ | 48 | 48 | $1.00\times10^{-2}$ | 3.96 | $7.92\times10^{-5}$ | steady |
| 10 | 10 | $10^6$ | 64 | 64 | $3.16\times10^{-2}$ | 1.04 | $2.07\times10^{-5}$ | steady |
| 10 | 100 | $10^6$ | 64 | 64 | $1.00\times10^{-1}$ | 0.23 | $4.51\times10^{-6}$ | steady |
| 10 | 1000 | $10^6$ | 64 | 64 | $3.16\times10^{-1}$ | 0.05 | $9.12\times10^{-7}$ | steady |
| 100 | 0.001 | $10^7$ | 48 | 48 | $1.00\times10^{-3}$ | 20.74 | $4.15\times10^{-4}$ | steady |
| 100 | 0.01 | $10^7$ | 96 | 96 | $3.16\times10^{-3}$ | 20.57 | $4.11\times10^{-4}$ | steady |
| 100 | 0.1 | $10^7$ | 96 | 96 | $1.00\times10^{-2}$ | 18.43 | $3.69\times10^{-4}$ | steady |
| 100 | 1 | $10^7$ | 48 | 48 | $3.16\times10^{-2}$ | 10.20 | $2.04\times10^{-4}$ | steady |
| 100 | 10 | $10^7$ | 96 | 96 | $1.00\times10^{-1}$ | 2.37 | $4.75\times10^{-5}$ | steady |
| 100 | 100 | $10^7$ | 96 | 96 | $3.16\times10^{-1}$ | 0.47 | $9.35\times10^{-6}$ | steady |
| 100 | 10000 | $10^7$ | 192 | 192 | 3.16 | 0.02 | $3.10\times10^{-7}$ | steady |
| 1000 | 0.01 | $10^8$ | 160 | 160 | | | unsteady | |
| 1000 | 0.1 | $10^8$ | 128 | 128 | | | unsteady | |
| 1000 | 1 | $10^8$ | 128 | 128 | $1.00\times10^{-1}$ | 43.64 | $8.73\times10^{-4}$ | steady |
| 1000 | 10 | $10^8$ | 128 | 128 | $3.16\times10^{-1}$ | 6.00 | $1.20\times10^{-4}$ | steady |
| 1000 | 100 | $10^8$ | 128 | 128 | 1.00 | 0.87 | $1.75\times10^{-5}$ | steady |
| 10000 | 0.01 | $10^9$ | 128 | 128 | | | unsteady | |
| 10000 | 0.1 | $10^9$ | 128 | 128 | | | unsteady | |
| 10000 | 1 | $10^9$ | 128 | 128 | $3.16\times10^{-1}$ | 218.25 | $4.36\times10^{-3}$ | steady |
| 10000 | 10 | $10^9$ | 128 | 128 | 10.0 | 10.36 | $2.07\times10^{-4}$ | steady |
| 10000 | 100 | $10^9$ | 128 | 128 | 3.16 | 1.55 | $3.10\times10^{-5}$ | steady |
| 100000 | 1 | $10^{10}$ | 64 | 64 | 1.00 | 420.48 | $8.41{\times}10^{-3}$ | steady |

Table A.2: Summary of all numerical simulations with $E = 10^{-5}$.

| $B$ | $S$ | $Ra_{\mathcal{H}}$ | $n_r$ | $l_{max}$ | $\frac{N}{2\Omega}$ | $Re$ | $Ro$ | State |
|---|---|---|---|---|---|---|---|---|
| 0.001 | 0.001 | $10^3$ | 96 | 96 | $1.00{\times}10^{-6}$ | 0.04 | $7.94{\times}10^{-8}$ | steady |
| 0.01 | 0.001 | $10^4$ | 96 | 96 | $3.16{\times}10^{-6}$ | 0.04 | $8.33{\times}10^{-8}$ | steady |
| 0.1 | 0.001 | $10^5$ | 96 | 96 | $1.00{\times}10^{-5}$ | 0.09 | $1.74{\times}10^{-7}$ | steady |
| 1 | 0.001 | $10^6$ | 96 | 96 | $3.16{\times}10^{-5}$ | 0.76 | $1.51{\times}10^{-6}$ | steady |
| 1 | 0.01 | $10^6$ | 96 | 96 | $1.00{\times}10^{-4}$ | 0.76 | $1.51{\times}10^{-6}$ | steady |
| 1 | 0.1 | $10^6$ | 96 | 96 | $3.16{\times}10^{-4}$ | 0.76 | $1.51{\times}10^{-6}$ | steady |
| 1 | 1 | $10^6$ | 96 | 96 | $1.00{\times}10^{-3}$ | 0.75 | $1.51{\times}10^{-6}$ | steady |
| 1 | 10 | $10^6$ | 96 | 96 | $3.16{\times}10^{-3}$ | 0.47 | $9.37{\times}10^{-7}$ | steady |
| 1 | 100 | $10^6$ | 96 | 96 | $1.00{\times}10^{-2}$ | 0.12 | $2.33{\times}10^{-7}$ | steady |
| 10 | 0.001 | $10^7$ | 96 | 96 | $1.00{\times}10^{-4}$ | 5.57 | $1.11{\times}10^{-5}$ | steady |
| 10 | 0.01 | $10^7$ | 96 | 96 | $3.16{\times}10^{-4}$ | 5.56 | $1.11{\times}10^{-5}$ | steady |
| 10 | 0.1 | $10^7$ | 96 | 96 | $1.00{\times}10^{-3}$ | 5.54 | $1.11{\times}10^{-5}$ | steady |
| 10 | 1 | $10^7$ | 96 | 96 | $3.16{\times}10^{-3}$ | 4.17 | $8.34{\times}10^{-6}$ | steady |
| 10 | 10 | $10^7$ | 96 | 96 | $1.00{\times}10^{-2}$ | 1.13 | $2.25{\times}10^{-6}$ | steady |
| 10 | 100 | $10^7$ | 192 | 192 | $3.16{\times}10^{-2}$ | 0.27 | $5.33{\times}10^{-7}$ | steady |
| 100 | 0.001 | $10^8$ | 128 | 128 | $3.16{\times}10^{-4}$ | 22.69 | $4.54{\times}10^{-5}$ | steady |
| 100 | 0.01 | $10^8$ | 128 | 128 | $1.00{\times}10^{-3}$ | 22.37 | $4.47{\times}10^{-5}$ | steady |
| 100 | 0.1 | $10^8$ | 128 | 128 | $3.16{\times}10^{-3}$ | 20.04 | $4.01{\times}10^{-5}$ | steady |
| 100 | 1 | $10^8$ | 96 | 96 | $1.00{\times}10^{-2}$ | 11.43 | $2.29{\times}10^{-5}$ | steady |
| 100 | 10 | $10^8$ | 96 | 96 | $3.16{\times}10^{-2}$ | 2.81 | $5.26{\times}10^{-7}$ | steady |
| 1000 | 0.1 | $10^9$ | 160 | 160 | | | | unsteady |
| 1000 | 10 | $10^9$ | 96 | 96 | $1.00\times10^{-1}$ | 9.91 | $1.98\times10^{-5}$ | steady |
| 1000 | 100 | $10^9$ | 224 | 224 | $3.16\times10^{-1}$ | 1.34 | $2.68\times10^{-6}$ | steady |

Table A.3: Summary of all numerical simulations with $E = 10^{-6}$.

| $E$ | $B$ | $S$ | $Ra_{\mathcal{H}}$ | $n_r$ | $l_{max}$ | $\frac{N}{2\Omega}$ | $Re$ | $Ro$ |
|---|---|---|---|---|---|---|---|---|
| $10^{-4}$ | 1 | 0.001 | 10000 | 48 | 48 | $3.16\times10^{-4}$ | 0.300872 | $0.601744\times10^{-4}$ |
| $10^{-4}$ | 1 | 1 | 10000 | 48 | 48 | $1.00\times10^{-2}$ | 0.295070 | $0.590139\times10^{-4}$ |
| $10^{-4}$ | 1 | 10 | 10000 | 48 | 48 | $3.16\times10^{-2}$ | 0.217627 | $0.435254\times10^{-4}$ |
| $10^{-4}$ | 1 | 100 | 10000 | 48 | 48 | $1.00\times10^{-1}$ | 0.064906 | $0.129813\times10^{-4}$ |
| $10^{-4}$ | 1 | 1000 | 10000 | 48 | 48 | $3.16\times10^{-1}$ | 0.013037 | $0.260743\times10^{-5}$ |
| $10^{-4}$ | 10 | 1 | 10000 | 48 | 48 | $3.16\times10^{-2}$ | 2.243120 | $0.448624\times10^{-3}$ |
| $10^{-4}$ | 10 | 10 | 10000 | 48 | 48 | $1.00\times10^{-2}$ | 1.056216 | $0.211243\times10^{-3}$ |
| $10^{-4}$ | 10 | 100 | 10000 | 48 | 48 | $3.16\times10^{-1}$ | 0.594569 | $0.118914\times10^{-3}$ |
| $10^{-4}$ | 100 | 1 | 10000 | 48 | 48 | $1.00\times10^{-1}$ | 11.922864 | $0.238457\times10^{-2}$ |
| $10^{-4}$ | 100 | 10 | 10000 | 48 | 48 | $3.16\times10^{-1}$ | 11.787036 | $0.235741\times10^{-2}$ |
| $10^{-4}$ | 100 | 100 | 10000 | 48 | 48 | 1.00 | 0.243352 | $0.486704\times10^{-4}$ |
| $10^{-4}$ | 100 | 1000 | 10000 | 48 | 48 | 3.16 | 0.041440 | $0.828798\times10^{-5}$ |

Table A.4: Summary of all numerical simulations with $E = 10^{-4}$ and shell aspect ratio $\eta = 0.01$.

## Appendix B. Scaling analysis figures

Example figures of the term balances in the vorticity and temperature equations for a few representative high and low S models. These figures are used to verify our scaling predictions (i.e. that we have used the correct length scales in various terms) and to justify only considering certain terms in the governing equation in the scaling analyses, as they make clear that the balances we consider are both applicable in our two $S$ regimes, appropriately scaled in our analysis and that our volume-averaged model diagnostics are appropriate (as we could have chosen other diagnostic outputs from the simulations).

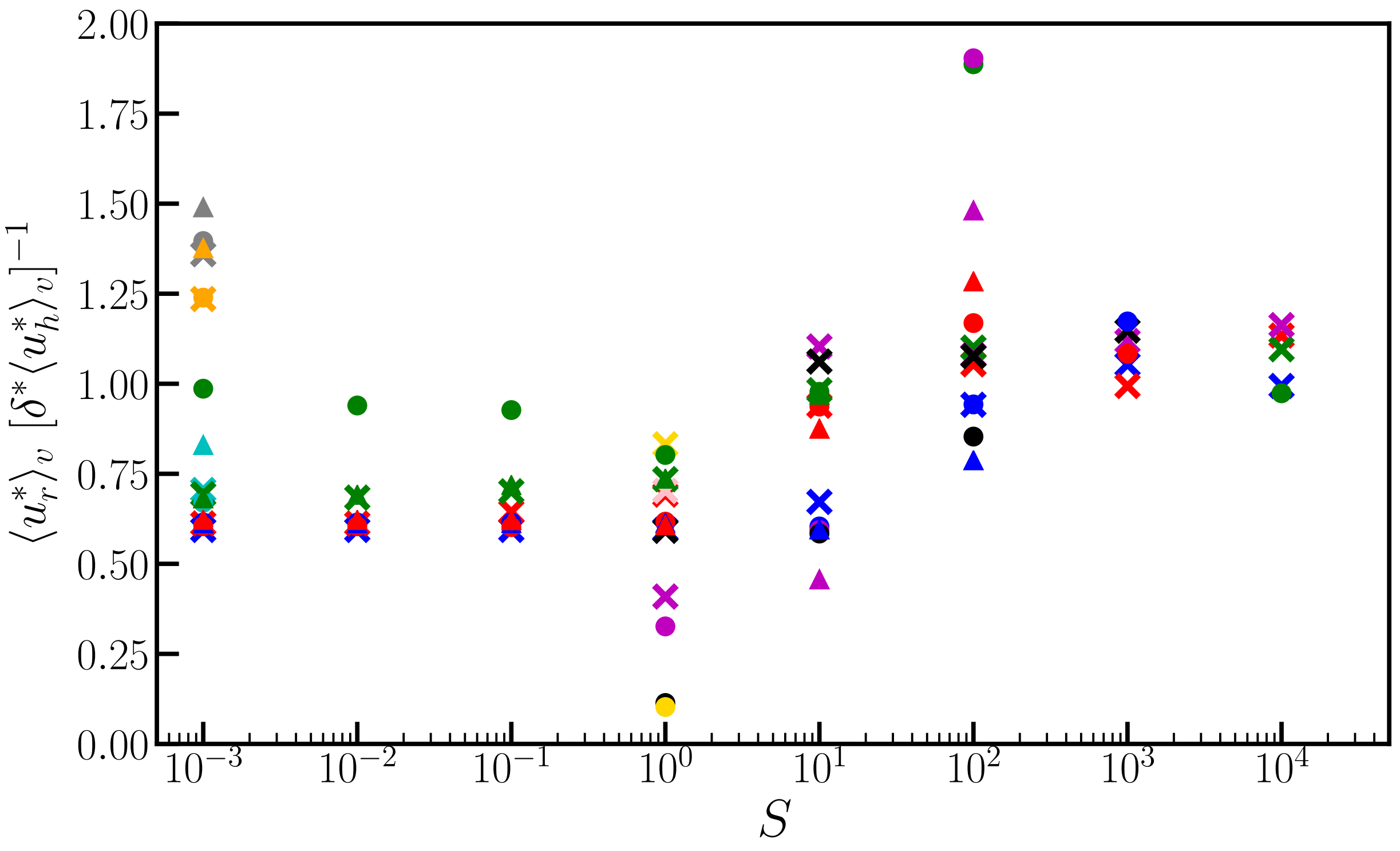

Figure B.1: Radial velocity scaled by $\delta^* \langle u_h^* \rangle_v$, where $\langle u_h^* \rangle_v$ is the average volume-averaged horizontal velocity, as a function of the stratification parameter, $S$, for all steady models. Symbol shapes represent the Ekman number, $E$, and colours represent the buoyancy parameter, $B$. The key is given in fig 6a.

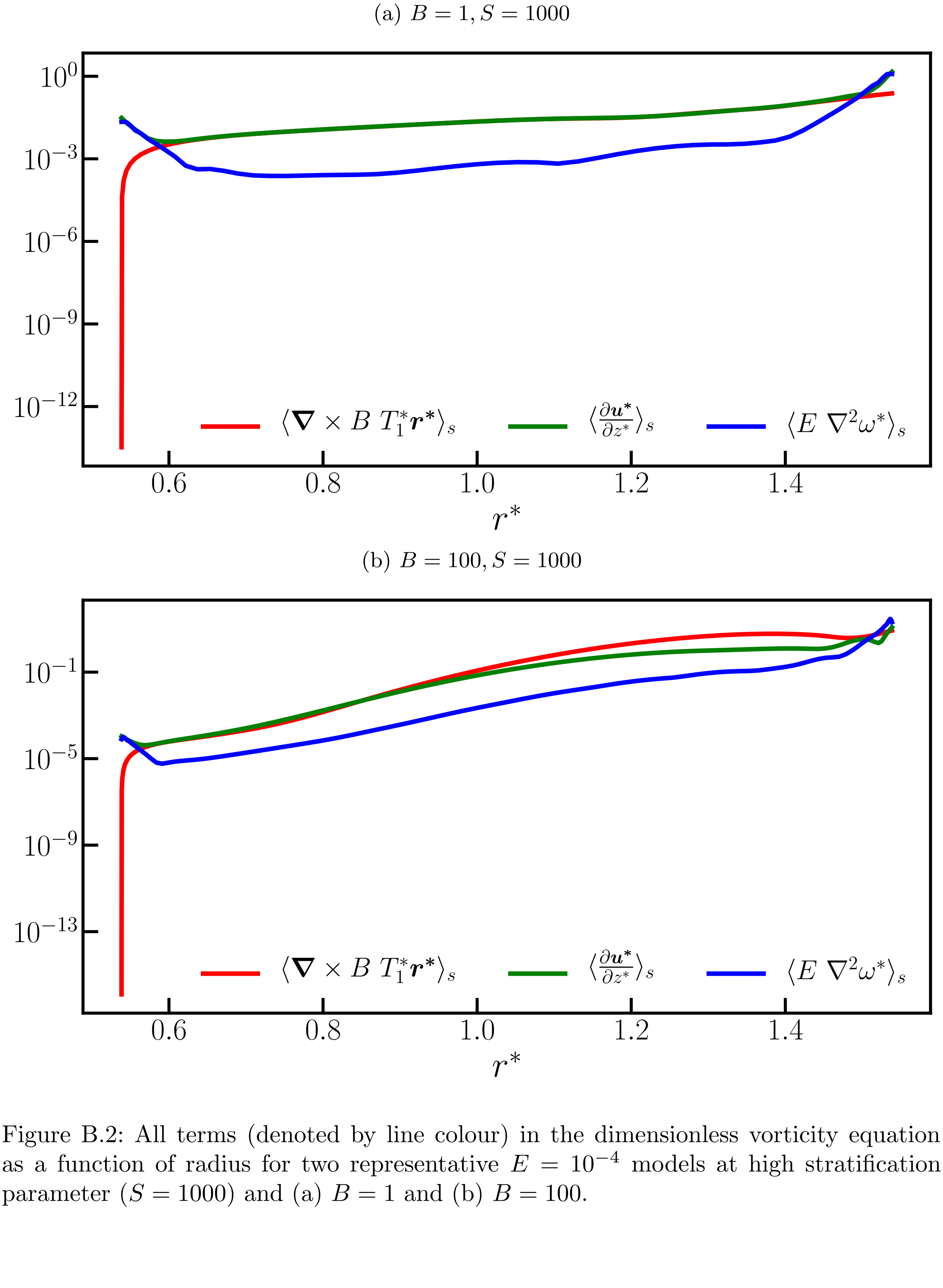

Figure B.2: All terms (denoted by line colour) in the dimensionless vorticity equation as a function of radius for two representative $E = 10^{-4}$ models at high stratification parameter ($S = 1000$) and (a) $B = 1$ and (b) $B = 100$.

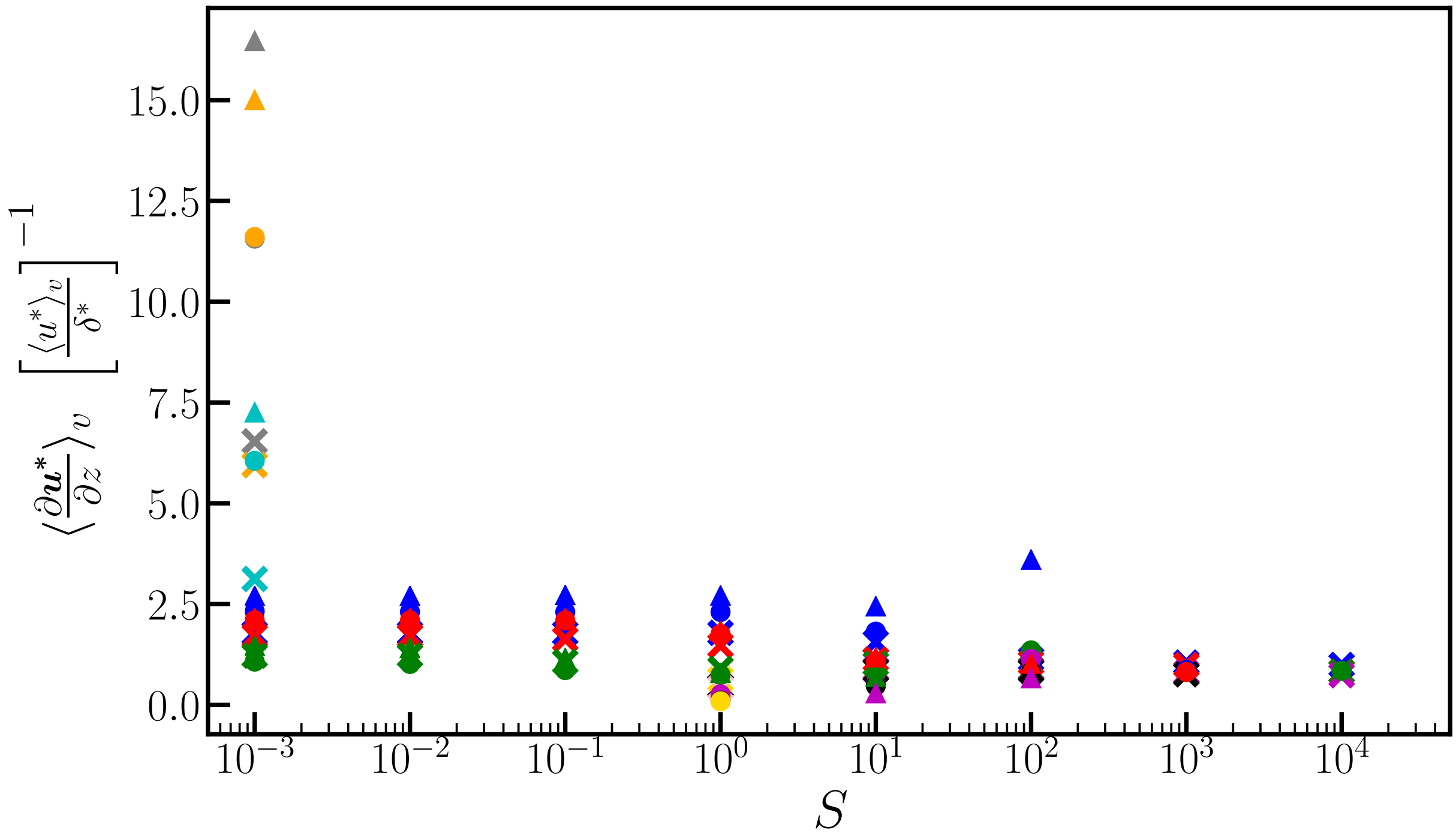

Figure B.3: Volume-averaged Coriolis term of the vorticity equation ($\frac{\partial \boldsymbol{u^*}}{\partial z^*}$), scaled by our approximation to that term ($\langle u^* \rangle_v / \delta^*$), as a function of the stratification parameter, $S$, for all steady models. Symbol shapes represent the Ekman number, $E$, and colours represent the buoyancy parameter, $B$. The key is given in fig 6a.

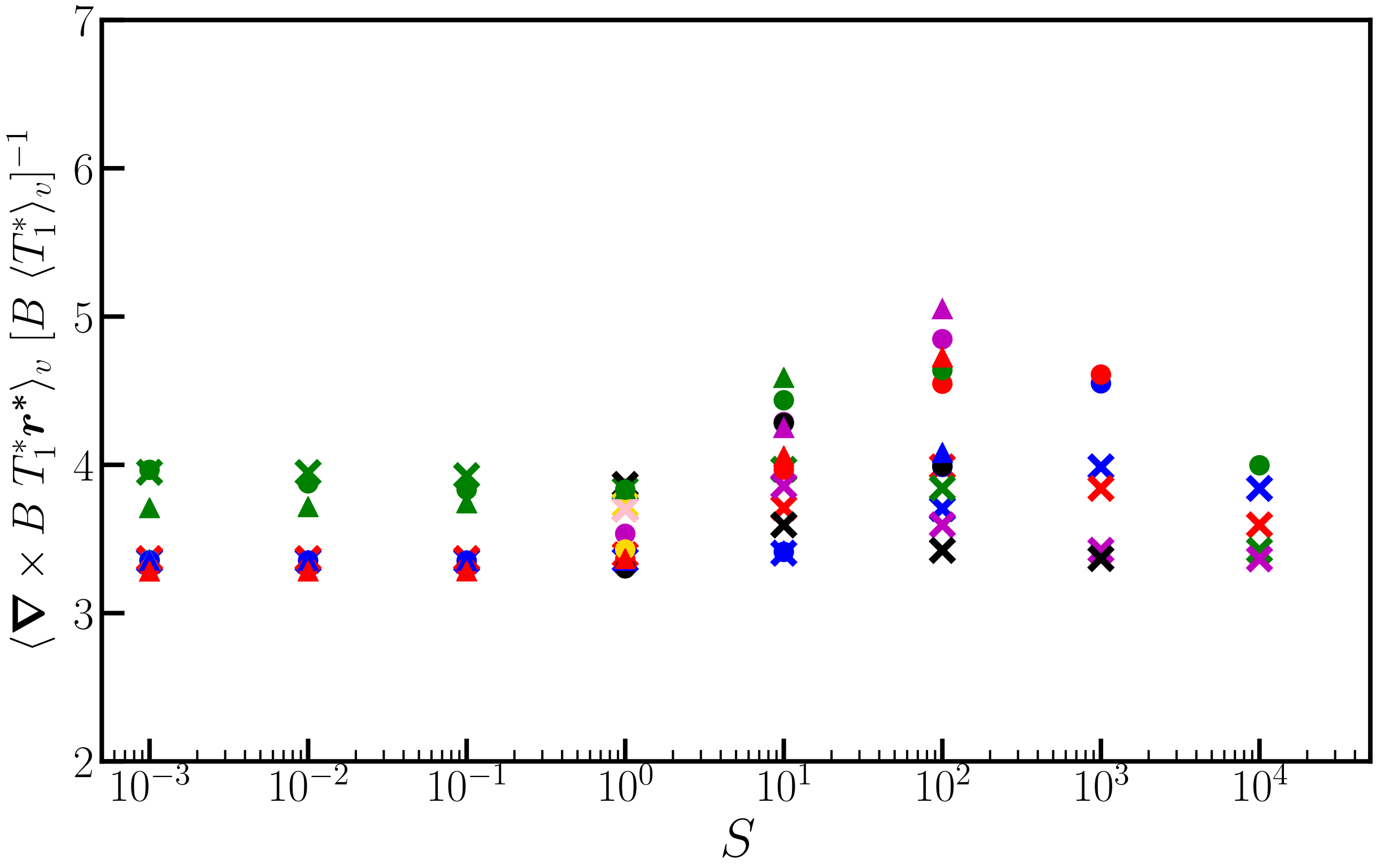

Figure B.4: Volume-averaged buoyancy term of the vorticity equation ($\boldsymbol{\nabla} \times B\ T_1^* \boldsymbol{r^*}$), scaled by our approximation to that term ($B\ \langle T_1^* \rangle_v$), as a function of the stratification parameter, $S$, for all steady models. Symbol shapes represent the Ekman number, $E$, and colours represent the buoyancy parameter, $B$. The key is given in fig 6a.

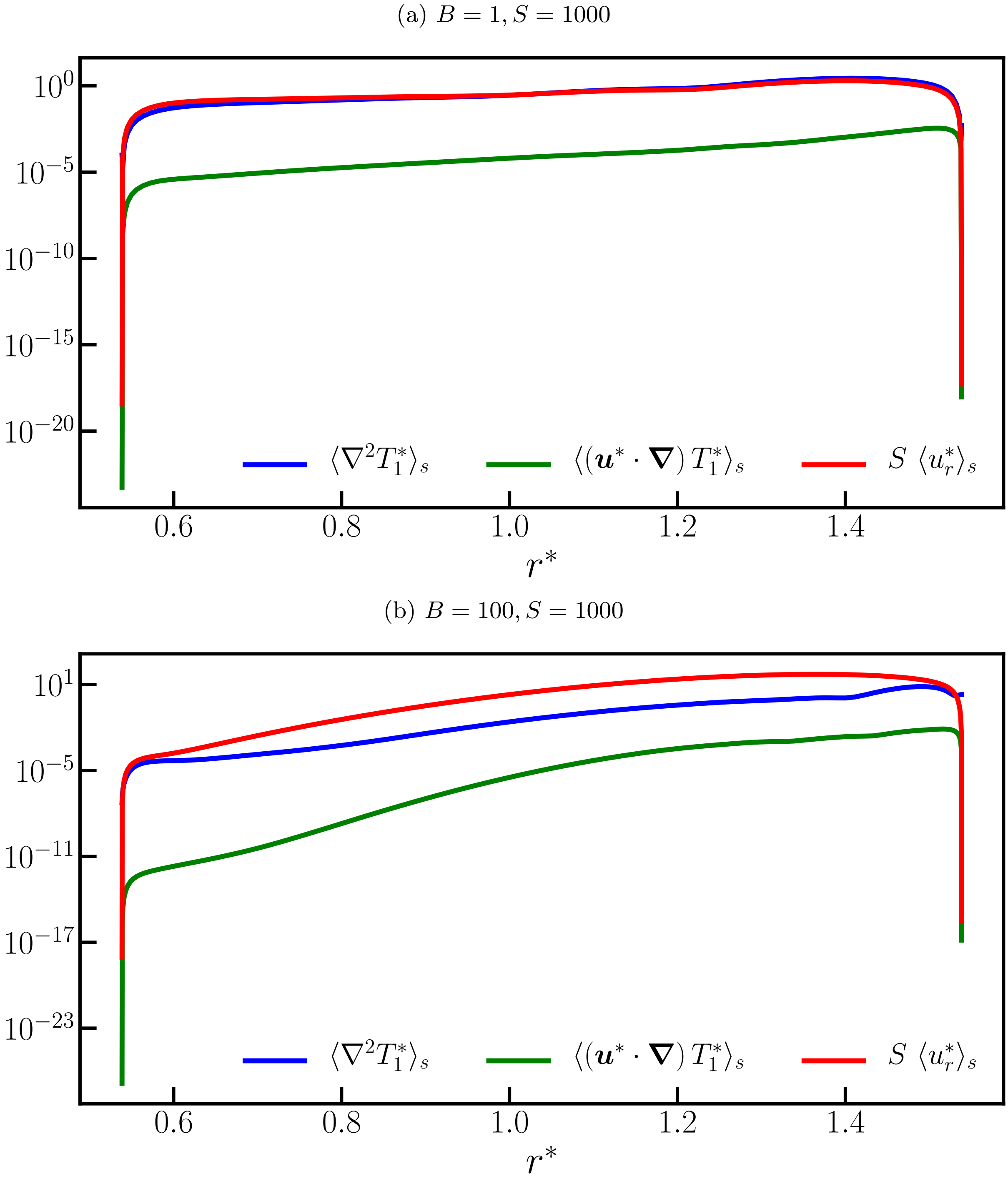

Figure B.5: All terms (denoted by line colour) in the dimensionless temperature equation as a function of radius for two representative $E = 10^{-4}$ models at high stratification parameter ($S = 1000$) and (a) $B = 1$ and (b) $B = 100$.

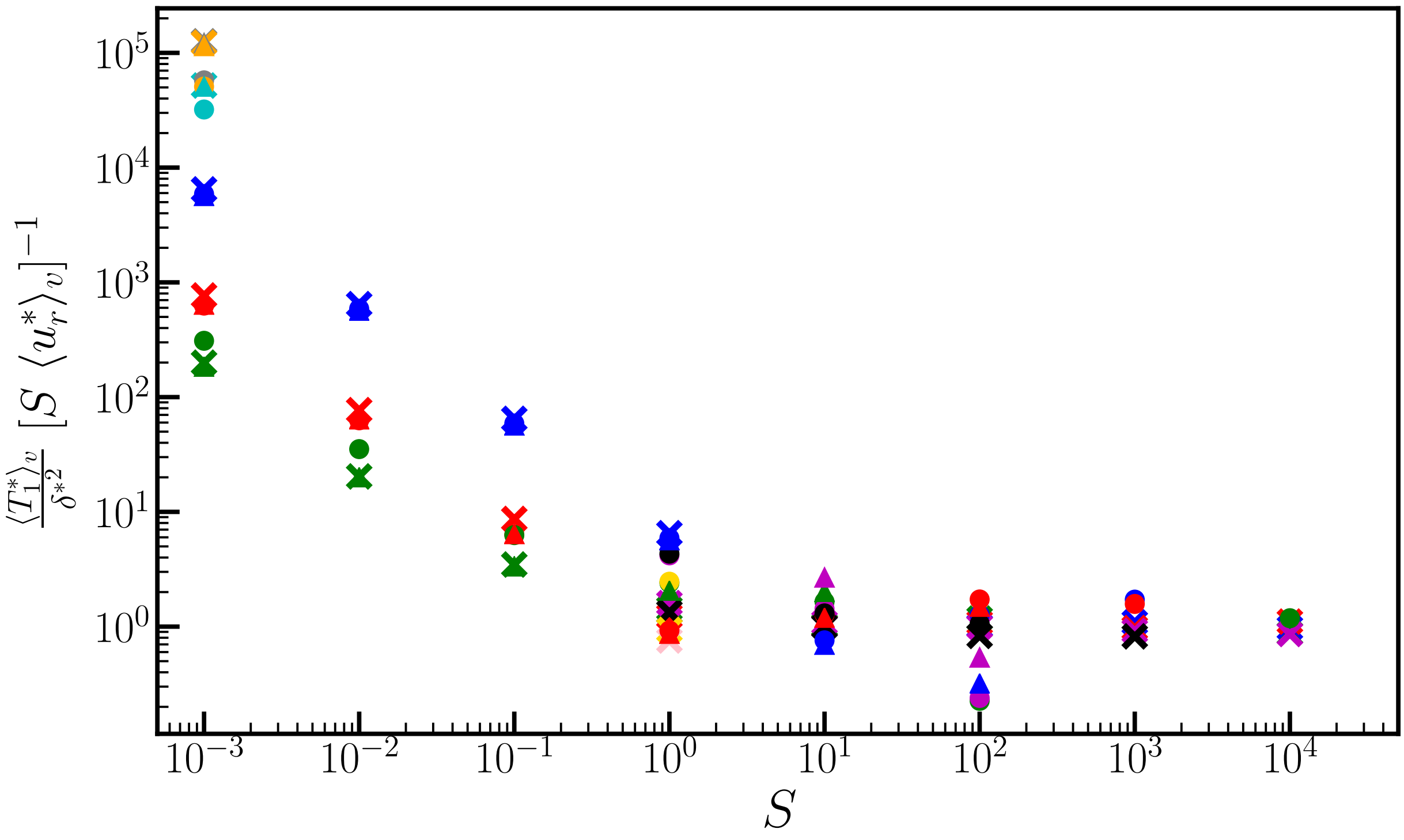

Figure B.6: Ratio of the two dominant terms in the temperature equation as a function of the stratification parameter, $S$, for all steady models. Symbol shapes represent the Ekman number, $E$, and colours represent the buoyancy parameter, $B$. The key is given in fig 6a.

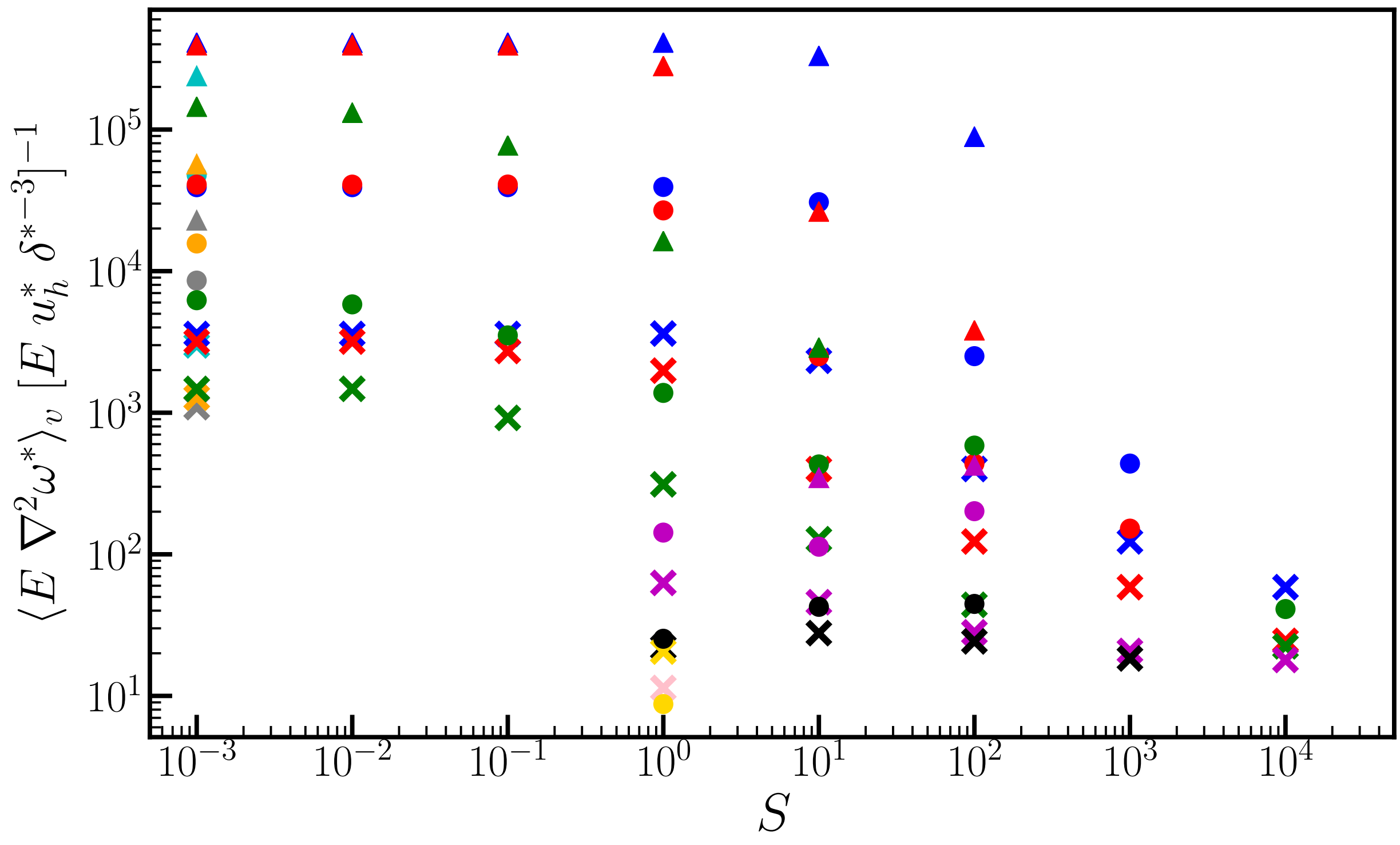

Figure B.7: Volume-averaged viscous term of the vorticity equation ($E \, \nabla^2 \omega^*$), scaled by the (incorrect) approximation to that term ($E \, u_h^* \delta^{*-3}$), as a function of the stratification parameter, $S$, for all steady models. Symbol shapes represent the Ekman number, $E$, and colours represent the buoyancy parameter, $B$. The key is given in fig 6a.

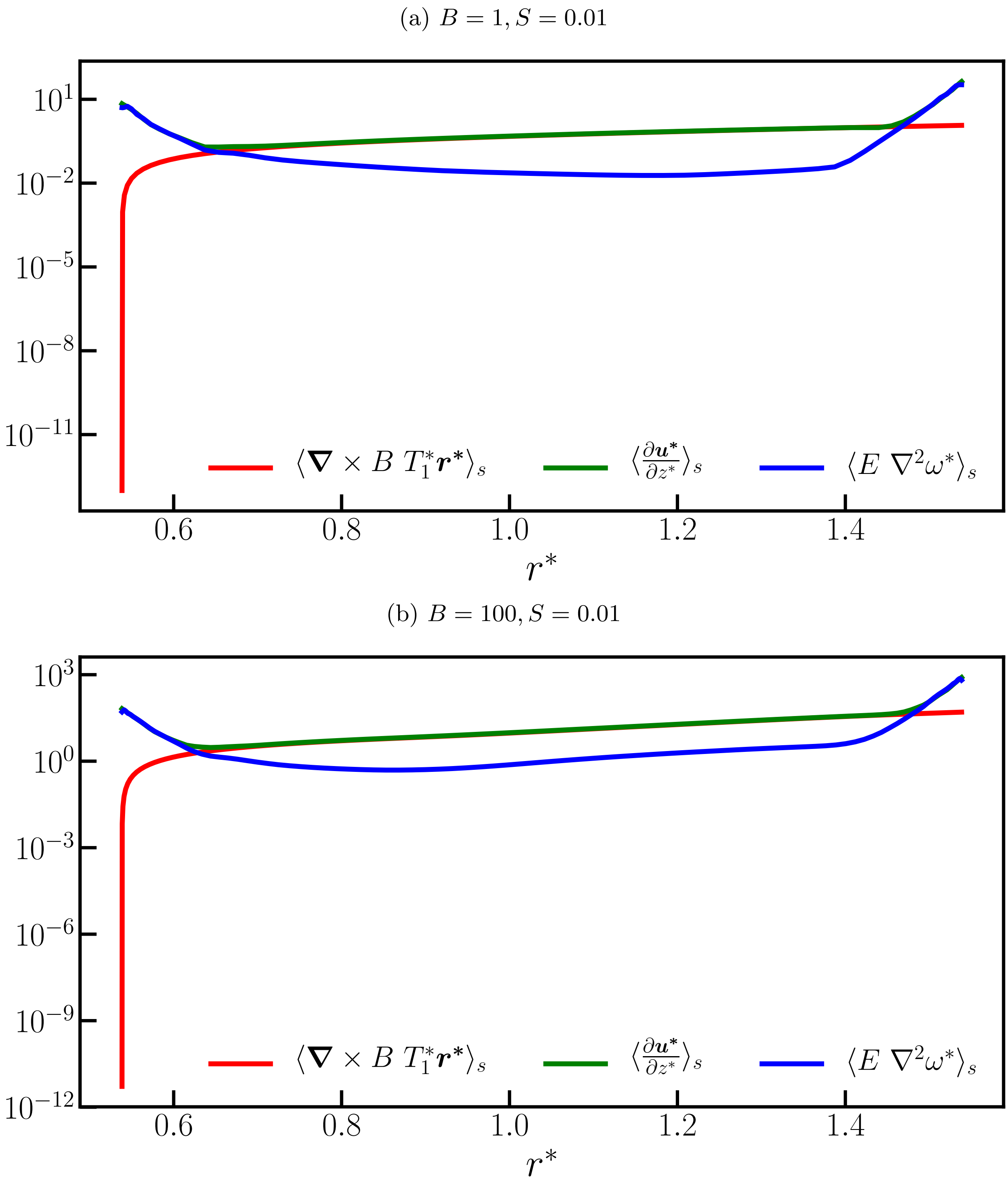

Figure B.8: All terms (denoted by line colour) in the dimensionless vorticity equation as a function of radius for two representative $E = 10^{-4}$ models at low stratification parameter ($S = 0.01$) and (a) $B = 1$ and (b) $B = 100$.

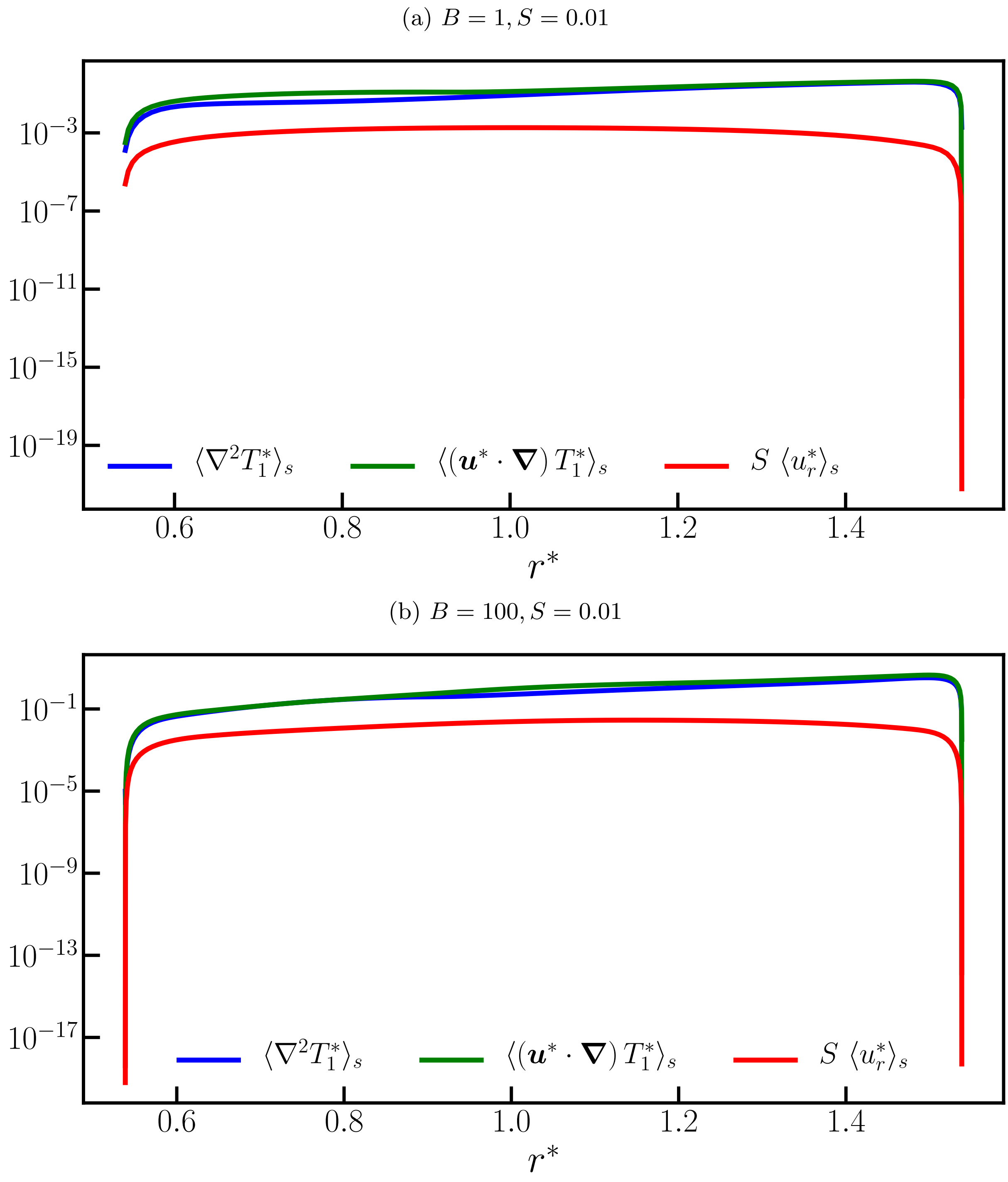

Figure B.9: All terms (denoted by line colour) in the dimensionless temperature equation as a function of radius for two representative $E = 10^{-4}$ models at low stratification parameter ($S = 0.01$) and (a) $B = 1$ and (b) $B = 100$.

## Appendix C. Low shell aspect ratio dynamics

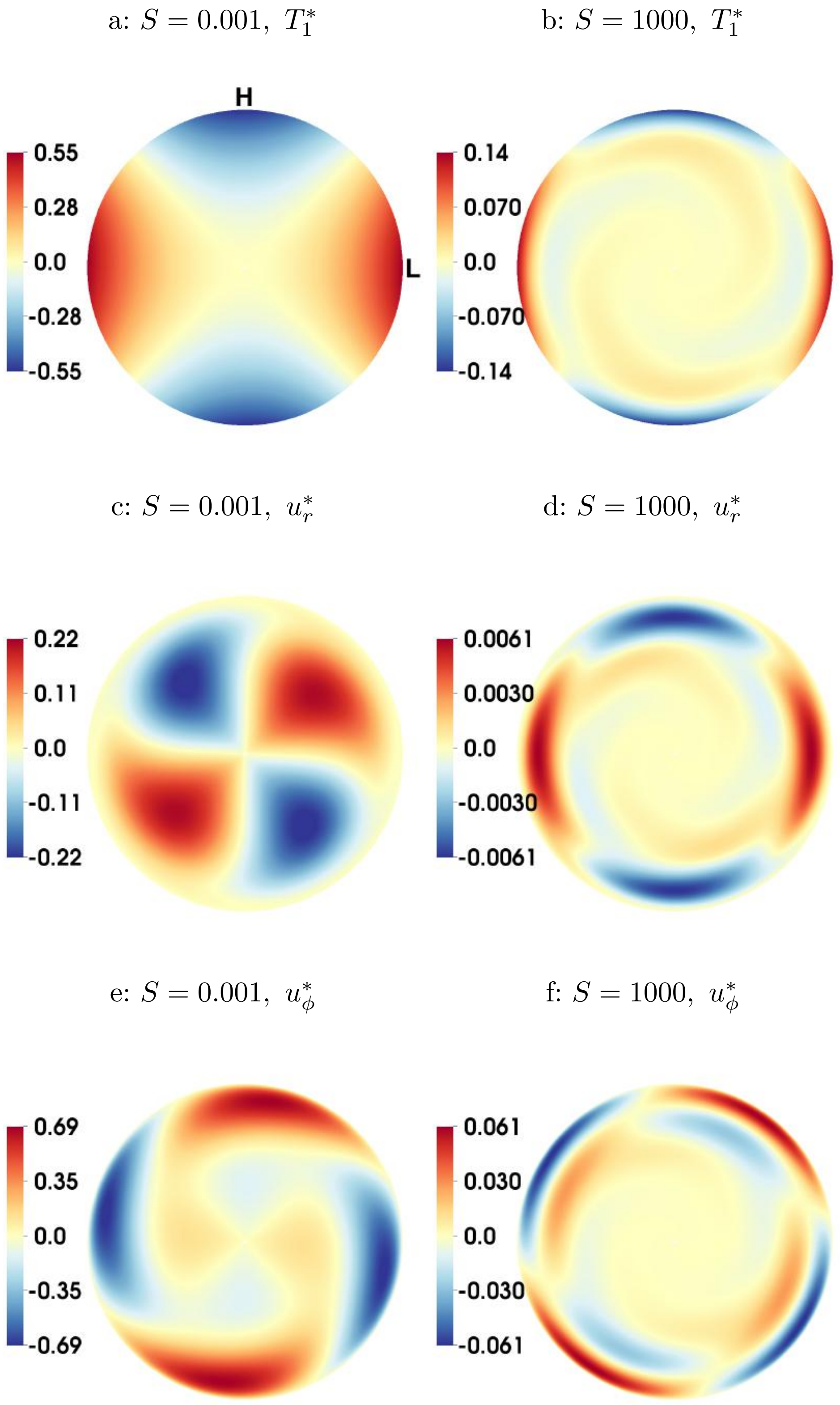

Figure C.1: Equatorial plots of $T_1^*$ (top), $u_r^*$ (middle) and $u_\phi^*$ (bottom) for models with shell aspect ratio $\eta = 0.01$ at $E = 10^{-4}$, $B = 1$ and $S = 0.001$ (left) and 1000 (right). Red indicates positive values and blue indicates negative values. Note the different colour scales. Locations of high (H) and low (L) outward heat flux are shown on the top left.